\documentclass[twocolumn]{aastex62}
\usepackage{subscript}
\usepackage{color}
\usepackage{subfigure}
 
\shorttitle{COLOR DISTRIBUTIONS OF GLOBULAR CLUSTER SYSTEMS}
\shortauthors{Lee, Chung \& Yoon}
\accepted{for publication in ApJS}

\begin{document}  
\title{N\MakeLowercase{ONLINEAR} C\MakeLowercase{OLOR}--M\MakeLowercase{ETALLICITY} R\MakeLowercase{ELATIONS OF} G\MakeLowercase{LOBULAR} C\MakeLowercase{LUSTERS}. VIII. R\MakeLowercase{EPRODUCING} C\MakeLowercase{OLOR} D\MakeLowercase{ISTRIBUTIONS OF} I\MakeLowercase{NDIVIDUAL} G\MakeLowercase{LOBULAR} C\MakeLowercase{LUSTER} S\MakeLowercase{YSTEMS IN THE} V\MakeLowercase{IRGO AND} F\MakeLowercase{ORNAX} G\MakeLowercase{ALAXY} C\MakeLowercase{LUSTERS}}  
\correspondingauthor{Suk-Jin Yoon}
\email{sjyoon0691@yonsei.ac.kr}
\author{Sang-Yoon Lee}
\affiliation{Center for Galaxy Evolution Research, Yonsei University, Seoul 03722, Republic of Korea}
\author{Chul Chung}
\affiliation{Center for Galaxy Evolution Research, Yonsei University, Seoul 03722, Republic of Korea}
\author{Suk-Jin Yoon}
\affiliation{Center for Galaxy Evolution Research, Yonsei University, Seoul 03722, Republic of Korea}
\affiliation{Department of Astronomy, Yonsei University, Seoul 03722, Republic of Korea}

\begin{abstract} 
The color distributions of globular clusters (GCs) in individual early-type galaxies show great diversity in their morphology.
Based on the conventional ``linear'' relationship between the colors and metallicities of GCs, the GC metallicity distributions inferred from colors and in turn their formation histories, should be as diverse as they appear.
In contrast, here we show that an alternative scenario rooted in the ``nonlinear'' nature of the metallicity-to-color transformation points to a simpler and more coherent picture.
Our simulations of the color distributions for $\sim$\,80 GC systems in early-type galaxies from the ACS Virgo and Fornax Cluster Surveys suggest that the majority ($\sim$\,70\,$\%$) of early-type galaxies have old ($\sim$\,13 Gyr) and coeval GCs. Their variety in the color distribution morphology stems mainly from one parameter, the mean metallicity of a GC system.
Furthermore, the color distributions of the remaining ($\sim$\,30\,$\%$) GC systems are also explained by the nonlinearity scenario, assuming additional young or intermediate-age GCs with a number fraction of $\sim$\,20\,$\%$ of underlying old GCs.
Our results reinforce the nonlinearity explanation for the GC color bimodality and provide a new perspective on early-type galaxy formation in the cluster environment, such as the Virgo and Fornax galaxy clusters.
\end{abstract} 
\keywords{galaxies: clusters: individual (Virgo, Fornax) -- galaxies: elliptical and lenticular, cD -- galaxies: evolution -- galaxies: star clusters: general -- globular clusters: general}

\section{I\MakeLowercase{ntroduction}}
\label{introduction}

Most galaxies harbor a system of globular clusters (GCs) that closely traces the formation history of its host galaxy.
GC systems predominantly exhibit bimodal color distribution functions (CDFs), and that phenomenon has been a major topic in the field of the extragalactic GC research~\citep[e.g.,][see also  \citealt{2004Natur.427...31W,2006ARA&A..44..193B} for reviews and references therein]{1993AJ....105.1762O,1993MNRAS.264..611Z,1999AJ....118.1526G,2001AJ....122.1251K,2001AJ....121.2974L,2006ApJ...639...95P,2008ApJ...674..857L,2009ApJS..180...54J,2010ApJ...710...51B,2011MNRAS.416..155F,2011MNRAS.413.2943F,2011MNRAS.415.3393F,2011ApJ...728..116L,2012MNRAS.420...37B,2012A&A...539A..54C,2012MNRAS.422.3591C,2012MNRAS.421..635F,2013ApJ...763...40K,2013MNRAS.436.1172U,2014A&A...564L...3C,2014MNRAS.437..273K,2016MNRAS.458..105K,2015A&A...574A..21R,2016ApJ...822...95C,2017ApJ...835..101H}.
Understanding the origin of the bimodal CDFs of extragalactic GC systems should offer valuable constraints on the evolutionary paths taken by their host galaxies.

The key assumption of GC formation scenarios in attempts to explain GC color bimodality is the existence of two GC subgroups with distinct mean metallicities.
These explanations invoke different origins for metal-poor and metal-rich subgroups, including (a) the metal-rich population is the product of major merging between two gas-rich spiral galaxies~\citep{1972ApJ...178..623T,1992ApJ...384...50A,1993MNRAS.264..611Z,1995AJ....109..960W,1997AJ....114.2381M}; (b) the lower-mass galaxies with metal-poor GCs are accreted onto a massive galaxy~\citep{1987ApJ...313..112M,1998ApJ...501..554C,2002ApJ...567..853C,1999A&AS..138...55H}; and (c) multiphase dissipational collapse leads to the discrete metallicity groups of GCs~\citep{1994ApJ...429..177H,1997AJ....113..887F,1999AJ....117..855H,santos03}.
A modern way to describe the formation of early-type galaxies and their GC systems, such as the two-phase formation scenario~\citep[][see also \citealt{2010ApJ...725.2312O}]{2011MNRAS.413.2943F,2013ApJ...773L..27P,2016ApJ...822...70L,2018Natur.555..483B}, is more in line with (b).
Current hierarchical galaxy formation models in the $\Lambda$CDM cosmology show that tens of thousands of small (proto-)galaxies have been involved in making one single galaxy. This is important because the great degree of complexity seems to leave little room for the existence of just two GC subpopulations in each galaxy.

An alternative explanation was proposed by \citet[hereafter Paper I]{2006Sci...311.1129Y}, in which the nonlinear metallicity-to-color conversion  creates GC color bimodality even from a broad, unimodal metallicity distribution function (MDF), without invoking two distinct GC subgroups within one galaxy.
Paper I incorporates a realistic treatment of core helium-burning, horizontal-branch stars in stellar population modeling, and find that nonlinearity is greatly enhanced by the inclusion of such stars.
\citet[Paper II]{2011ApJ...743..149Y} and \citet[Paper IV]{2013ApJ...768..137Y} demonstrated that the GC CDFs vary systematically for GC samples in M87 (Paper II) and M84 (Paper IV), respectively.
Using their theoretical color--metallicity relations (CMRs), they reproduced the CDF morphologies with different filter combinations ($u-g$, $u-z$, and $g-z$), which are in good agreement with the observations.
\citet[Paper III]{2011ApJ...743..150Y} demonstrated that by applying nonlinear color-to-metallicity conversions, the inferred GC MDFs have skewed, broad shapes and are remarkably similar to the MDF shapes of halo field stars in their host galaxies, implying common evolutionary histories between GCs and halo stars.
\citet[Paper V]{2013ApJ...768..138K} and \citet[Paper VII]{2017ApJ...843...43K} showed that the diverse morphology of absorption-line index (e.g., H$\beta$ and Mg2) distributions for M31 GCs (Paper V) and NGC 5128 GCs (Paper VII) can readily be reproduced by nonlinear ``index''--metallicity relations, exactly analogous to the nonlinear CMRs.
\citet[Paper VI]{2016ApJ...818..201C} further extended these results to the infrared Calcium II Triplet index (CaT) and proposed the nonlinear CaT--metallicity relation as the origin of observed bimodal CaT distributions of GCs in 12 early-type galaxies.

In this paper of the series, we attempt to reproduce quantitatively color distributions of individual GC systems in the ACS Virgo Cluster Survey~\citep[ACSVCS;][]{2004ApJS..153..223C} and ACS Fornax Cluster Survey~\citep[ACSFCS;][]{2007ApJS..169..213J}.
Our model shows that the majority ($\sim$\,$70\,\%$) of early-type galaxies have old ($\sim$\,13 Gyr) and coeval GCs.
The remaining ($\sim$\,30\,$\%$) galaxies are readily reproduced by including additional intermediate-age GCs.
The paper is organized as follows.
Section 2 describes the observational data set that we examined.
Section 3 presents our simulated CDFs of individual GC systems and derives the best-fit parameters through the Kolmogorov--Smirnov (K-S) analysis.
In Section 4, we examine the derived parameters for the GC systems as functions of host galaxy luminosity.
Section 5 investigates the GC systems whose derived ages are younger than the majority.
Our interpretation of the GC CDFs from the viewpoint of galaxy formation is given in Section 6.
Finally, we conclude in Section 7.

\begin{figure*}
\begin{center}
\includegraphics[width=14.0cm]{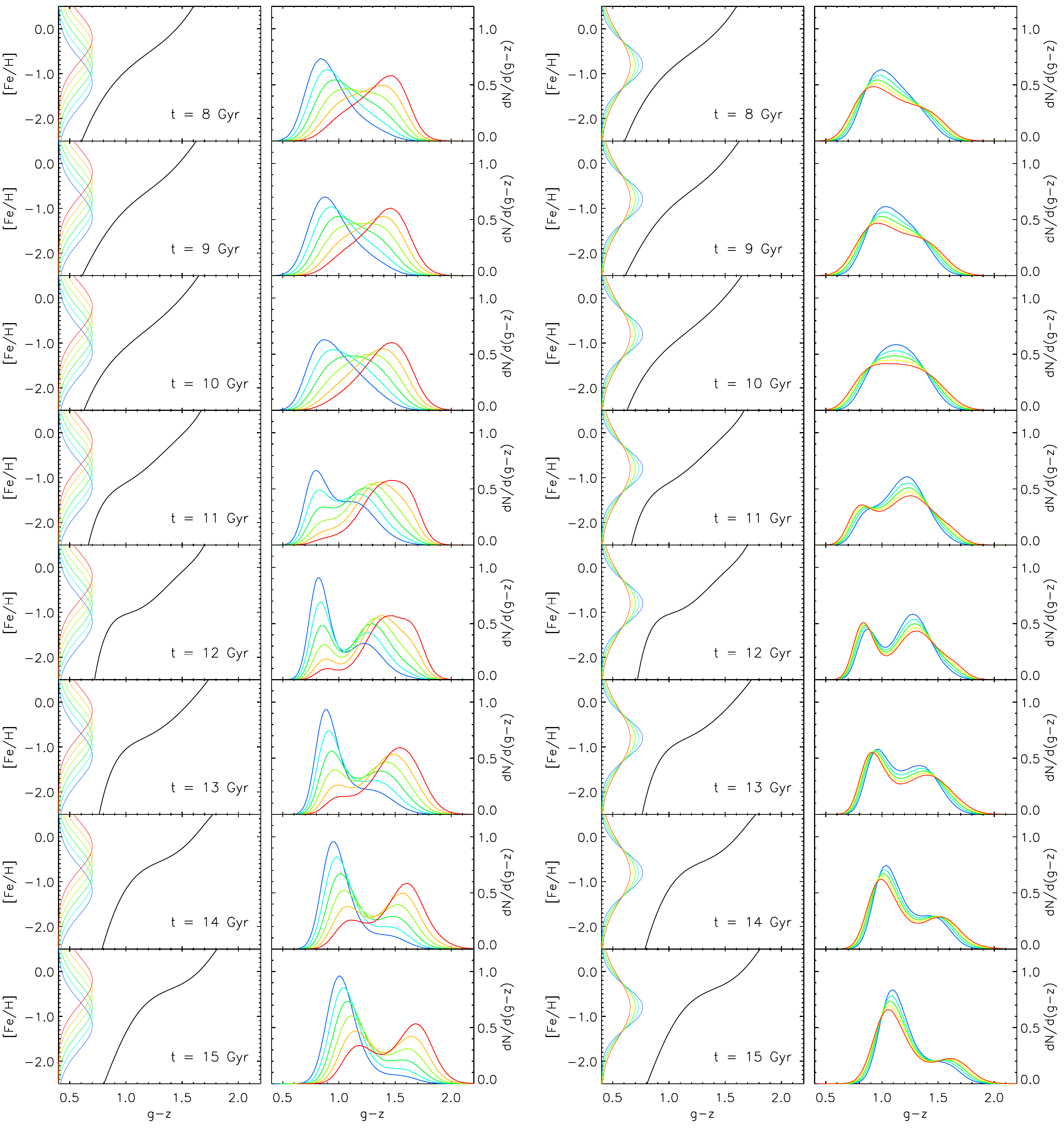}   
\caption{Monte Carlo simulations of $g-z$ color distributions for $10^6$ model GCs. 
Left two columns: the simulated GC CDFs according to the change of the mean metallicity and age.
In the left column, the $g-z$ CMRs (black solid lines) and the input MDFs are shown.
The mean metallicity spans from ${\langle{\rm}}$[Fe/H]${\rangle}=-1.2$ (blue) to ${\langle{\rm}}$[Fe/H]${\rangle}=-0.2$ (red) with equal [Fe/H] intervals of 0.2 dex.
The dispersion of MDFs is taken as a fixed value, $\sigma($[Fe/H]$)=0.55$. 
In the right column, the same color code is used for the histogram of the CDFs.
Right two columns: the simulated GC CDFs according to the change of the $\sigma$([Fe/H]) and age.
In the left column, the mean metallicity of the MDFs is taken as a fixed value, ${\langle{\rm}}$[Fe/H]${\rangle}=-0.8$.
The dispersion of the MDFs spans from $\sigma($[Fe/H]$)=0.45$ (blue) to $\sigma($[Fe/H]$)=0.65$ (red) by equal $\sigma$([Fe/H]) intervals of 0.05 dex.
In the right column, the same color code is used for the histogram of the CDFs.}
\label{fig1}
\end{center}
\end{figure*}

\section{D\MakeLowercase{ATA}}
\label{data}
The data used herein are from the ACSVCS~\citep{2009ApJS..180...54J} and ACSFCS~\citep{2015ApJS..221...13J} GC catalogs, which obtained GC photometry for a total of 143 early-type galaxies in the F475W and F850LP filters (hereafter $g$ and $z$) using \textit{HST}\slash ACS imaging.
The ACSVCS and ACSFCS provide the deepest and most homogeneous photometric catalogs of extragalactic GC systems so far.
The galaxies span a wide luminosity range ($-15.1<M_{B}<-22.3$).
The morphologies of the GC CDFs show great diversity.
To avoid the small number statistics, we examine the galaxies with a number of observed GCs,\footnote{We use the GC candidates with $p\textsubscript{GC}\geq0.5$, where $p\textsubscript{GC}$ is a probability that a source is an actual GC. See Section 7 in \citet{2009ApJS..180...54J} for more details.} $N\textsubscript{GC}$, greater than 50, and our sample consists of 78 galaxies (56 and 22 galaxies in the Virgo and Fornax clusters, respectively).
A caveat is that the field of view ($202\arcsec\times202\arcsec$) of the ACS\slash WFC only takes in the GCs of the inner region for large galaxies, and thus the GC lists for large galaxies do not fully represent their entire populations of GCs above the detection limit.

\section{R\MakeLowercase{eproducing} CDF\MakeLowercase{s} \MakeLowercase{of} I\MakeLowercase{ndividual} GC S\MakeLowercase{ystems}}
\label{reproducingcolordistributionfunctionsofindividualglobularclustersystems}

\subsection{Modeling of GC Color Distributions}
\label{modelingofglobularclustercolordistributions}

Our main goal is to test the nonlinear-CMR scenario for the GC color bimodality of early-type galaxies.
The theoretical $g-z$ CMRs are based on the Yonsei Evolutionary Population Synthesis (YEPS) model.
The YEPS model is described in detail in \citet{2013ApJS..204....3C,2017ApJ...842...91C}.
To simulate GC CDFs, we generate the $g-z$ color distributions of $10^{6}$ model GCs with various ages, mean {[Fe\slash H]} $({\langle{\rm}}${[Fe\slash H]}${\rangle})$, and dispersion $(\sigma$({[Fe\slash H]})).
The metallicity spread of a GC system is assumed to be a Gaussian normal distribution.
For the transformation from MDFs to CDFs, we use the fifth-order polynomial fit to the model data.
For a realistic comparison, we apply the photometric error based on the observed magnitude--error relations and luminosity functions~\citep{2010ApJ...717..603V} of the GC system of interest.
Figure~\ref{fig1} presents our CDF models according to the change of the input parameters.
Table~\ref{table:table1} summarizes the input parameters of our simulations.

\begin{deluxetable*}{lcc}
\tablecaption{Input Parameters of Simulated Color Distribution Functions \label{table:table1}}
\tablecolumns{3}
\tablenum{1}
\tablewidth{0pt}
\tablehead{
\colhead{Parameter} &
\colhead{Range} &
\colhead{Grid Interval}
}
\startdata
The age of a GC system, $t$ (Gyr) &8.0 $\sim$ 15.0 &0.1\\
The mean [Fe/H] of a GC system, ${\langle{\rm}}$[Fe/H]${\rangle_{GC}}$ (dex) & $-$1.80 $\sim$ 0.80 &0.05\\
The dispersion of [Fe/H] distribution, $\sigma$([Fe/H]) (dex) &0.45 $\sim$ 0.65 &0.05\\
\enddata
\end{deluxetable*}

\begin{figure*}
\begin{center}
\includegraphics[width=13.0cm]{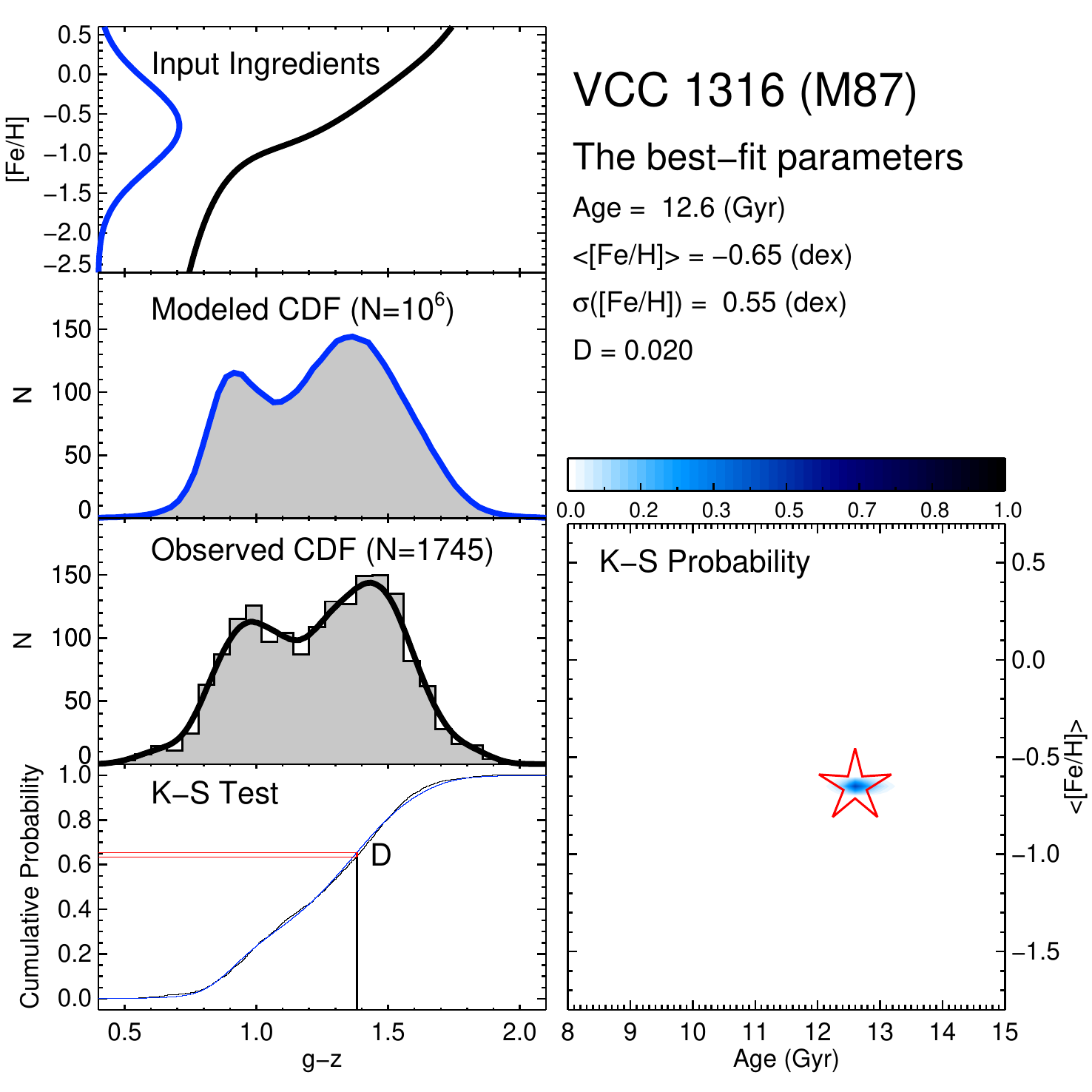}   
\caption{Monte Carlo simulation of the $g-z$ color distribution for the case of the VCC 1316 (M87) GC system. Top left panel: the ($g-z$)--[Fe/H] relation and assumed GC metallicity distributions of our best model. Second left panel: the $g-z$ color distribution of $10^6$ model GCs, which is transformed from the assumed MDF using the theoretical relation shown in the left top panel. Third left panel: the observed $g-z$ color histogram for 1745 GCs belonging to VCC 1316. The black line is a smoothed histogram with a Gaussian kernel with $\sigma(g-z)=0.06$. Bottom left panel: the K-S test for the model CDF against the observed CDF. The blue and black curves represent the cumulative function of the best modeled and the observed CDFs, respectively. The difference between the red lines shows the K-S statistic $D$. Right panel: the $p$-value contour plot of the K-S probability in the age--[Fe/H] plane with the color scale from white to black for increasing K-S probability. The red star marks the location of the highest probability.}
\label{fig2}
\end{center}
\end{figure*}

\subsection{The Kolmogorov--Smirnov Test}
\label{thekolmogorov-smirnovtest}

To determine the best-fit parameters, the two-sample K-S test is applied to the observed and modeled CDFs within the parameter space of age, ${\langle{\rm}}${[Fe\slash H]}${\rangle}$, and $\sigma$({[Fe\slash H]}) (Table 1).
Figure~\ref{fig2} shows an example of the best model CDF, the observed CDF, and the K-S statistics for the case of VCC 1316 (M87) that has the largest GC sample.
The right contour plot shows the $p$-value from the K-S statistics when $\sigma(${[Fe\slash H]}$)=0.55$ dex.
One can take a probability of $p>0.05$ for the observed and modeled CDFs being drawn from the same underlying distributions.

\subsection{Four Different Types of CDFs}
\label{fourdifferenttypesofcolordistributionfunctions}

Figure~\ref{fig3} presents our best model CDFs and the observed CDFs for the whole galaxy sample (78 ACSVCS\slash FCS galaxies) in order of $N\textsubscript{GC}$.
With the exception of FCC 21, the \emph{p}-values of the best-fit models for all the GC systems are greater than 0.05, implying the modeled CDFs reproduce the observed CDFs successfully.
Table~\ref{table2} presents the GC simulation result, along with the basic information on the host galaxies.

A closer look at Figure 3 suggests that the GC CDFs can be classified into four different types based on two factors.
The first factor is the number ratio of the blue and red GCs, which have a large effect on the overall shape of the CDFs.
To derive the number ratio of the blue and red GCs, we apply the Gaussian Mixture Modeling (GMM) code by \citet{2010ApJ...718.1266M}.
We classify the CDFs into Types 1 and 2 using the red GC fraction ($f\textsubscript{red}$) in the $g-z$ color distributions, and $f\textsubscript{red}=0.3$ divides the entire galaxy sample into halves.
The second factor is a model-derived parameter, the best-fit age, which enables more detailed classification of the CDFs.
The best-fit age used for dividing the subgroups, $a$ and $b$, is 11 Gyr.
We suspect that the galaxies classified as the subgroup $b$ (i.e., $t\textsubscript{best-fit}<11$ Gyr) possess additional intermediate-age GCs, weakening the bimodality made by the underlying old GCs.
Table~\ref{table:table3} gives the classifying criteria and the mean values of the model-derived parameters for each type.

Figure~\ref{fig4} shows the representative GC systems for the four GC CDF types (Types 1a, 1b, 2a, and 2b).
Type 1a is the most common GC CDF type and has a clear bimodal distribution with well-developed two peaks.
Most luminous galaxies ($M_{B}\lesssim-19$) favor this type, and VCC 1316 (M87) and VCC 1226 (M49) show the representative GC CDFs of this type.
For the case of Type 1b, two distinct peaks are not prominent. Compared to Type 1a, the distance between two peaks is shorter and the dip is less clear and shallower.
VCC 798 and VCC 1632 show the typical GC CDF shapes of this type.
This type is also found preferentially in luminous ($M_{B}\lesssim-19$) galaxies.
On the other hand, Type 2a is the second most common GC CDF type and has a dominant blue peak along with a broad red mound (e.g., VCC 1297) or a red tail (e.g., VCC 1303).
This type is typically found in less luminous ($M_{B}\gtrsim-19$) galaxies.
For the case of Type 2b, their CDFs have a blue peak but do not have additional structures such as a red mound or a red tail as seen in Type 2a.
The blue peak is less cuspy than Type 2a.
Thus, the overall shape of this type is a fat, skewed unimodal distribution.
This type is also found preferentially in less luminous ($M_{B}\gtrsim-19$) galaxies.

The lower part of Figure 4 shows the \emph{p}-value contours for the four types in the age--${\langle{\rm}}${[Fe\slash H]}${\rangle}$ plane for different assumptions of $\sigma(${[Fe\slash H]}$)$.
The contour plots for the whole galaxy sample are given in Figure 7 in the Appendix.
The contours of the K-S statistics are often stretched along the age axis, with narrow width along the {[Fe\slash H]}-axis.
This indicates that our K-S test is able to place a stronger constraint on the mean metallicity of given GC systems than the age.
From the \emph{p}-value contours, one can find two important elements that pin down the age of a GC system.
The first element is the strength of the bimodality of the color distribution.
Our GC CDF model predicts that the positions of two peaks and a dip are highly sensitive to the age of the GC system (see Figure 1).
The dip position in color corresponds to the quasi-inflection point along the CMR (Paper I) and becomes redder with increasing age.
As a result, the K-S statistics better determine the age if an observed CDF has clear bimodality and a well-shaped dip.
The other element is the total number of observed GCs.
The larger the sample sizes, the more compact the \emph{p}-value contours.
The Type 1a galaxies with high luminosity are well suited to these conditions and exhibit compact \emph{p}-value contours compared to other types.

When it comes to Type 2a, the sharp blue peak is one of the most distinguishing features of this type.
Since the sharp blue peak is generated by the steep slope of old CMRs, the contour converges to an old age (11\,$\sim$\,15 Gyr).
Some Type 2a CDFs that have only a feeble red tail show diagonal contours in the age range of 12\,$\sim$\,14 Gyr (e.g., VCC 1025, VCC 1475, VCC 1431, VCC 1303, and FCC 170), undergoing the age--metallicity degeneracy (see Figure 7 in the Appendix).
Weak red tails occur when the mean metallicity is so low (${\langle{\rm}}${[Fe\slash H]}${\rangle}<-1.3$) that only a few red GCs are present. In such cases, the quasi-inflection point of the CMR hardly affects the CDF shapes and thus accuracy in age dating becomes low. 
For instance, the 12 Gyr, ${\langle{\rm}}${[Fe\slash H]}${\rangle}=-1.3$ model and the 14 Gyr, ${\langle{\rm}}${[Fe\slash H]}${\rangle}=-1.8$ model have similar GC CDFs, with a sharp blue peak with a feeble red tail.
This uncertainty in age, however, does not affect our type classification.
Note also that Type 2a galaxies are relatively less luminous galaxies and have a small number of GCs, thus the confidence contours are rather broad.

\begin{figure*}
\includegraphics[width=16.0cm]{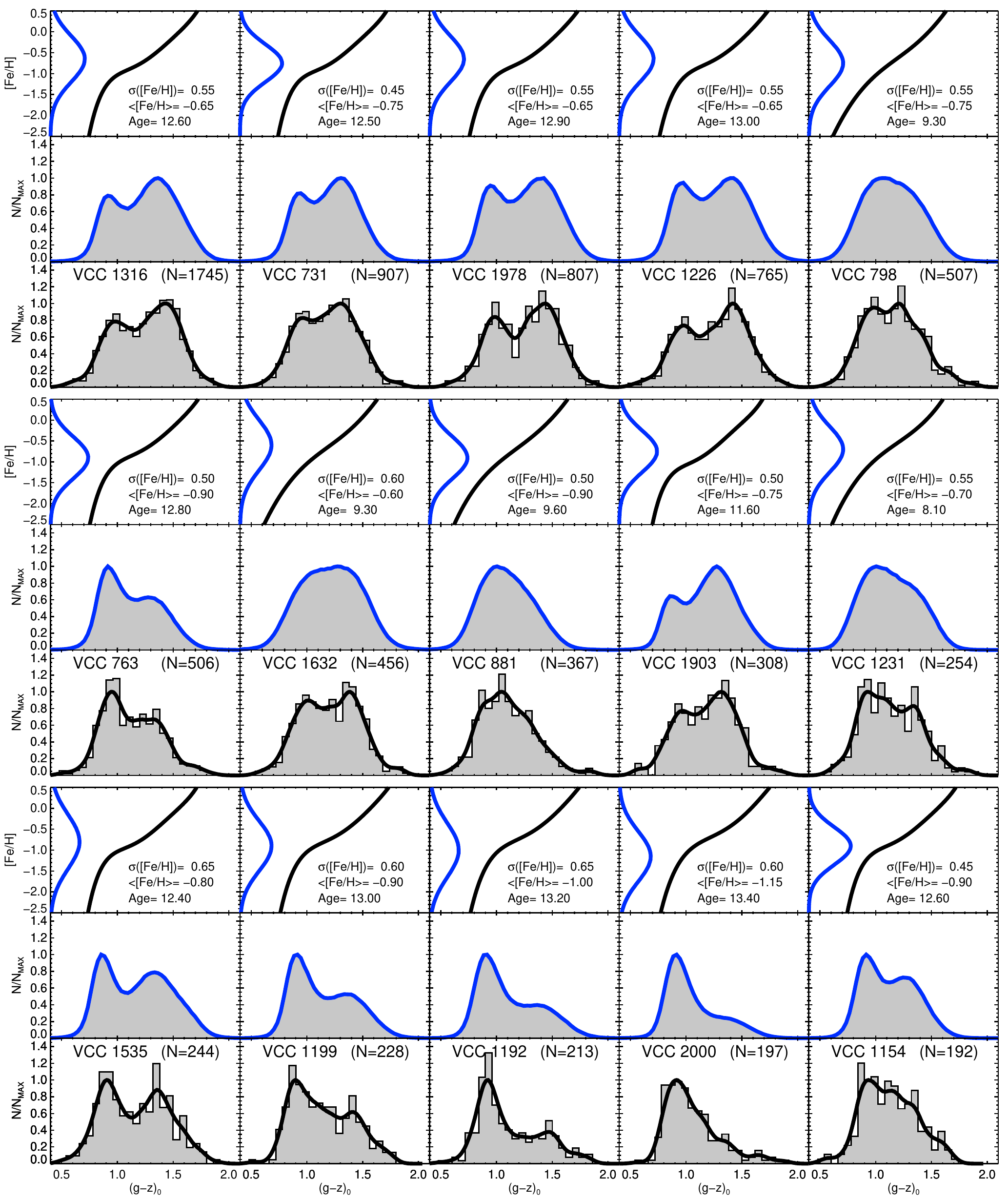}
\caption{The comparison of simulated $g-z$ distributions of GCs with observations for 78 ACSVCS and ACSFCS galaxies with $N\textsubscript{GC}>50$.
First, fourth, and seventh rows: the color-to-metallicity relations and metallicity distribution functions.
The black lines and blue curves show the ($g-z$)--[Fe/H] relation and assumed GC metallicity distributions of our best model, respectively.
The best-fit age, mean [Fe/H], and dispersion of MDFs are denoted.
Second, fifth, and eighth rows: the model CDFs.
The blue curves show the $g-z$ CDFs of our best models for individual galaxies. 
Third, sixth, and ninth rows: the observed CDFs.
The Black curves show the Gaussian kernel density estimations with $\sigma(g-z)=0.06$ for gray histograms.
The VCC name and the total number of observed GCs are denoted. 
}
\label{fig3}
\end{figure*}
\begin{figure*}
	\figurenum{3}
	\includegraphics[width=16.0cm]{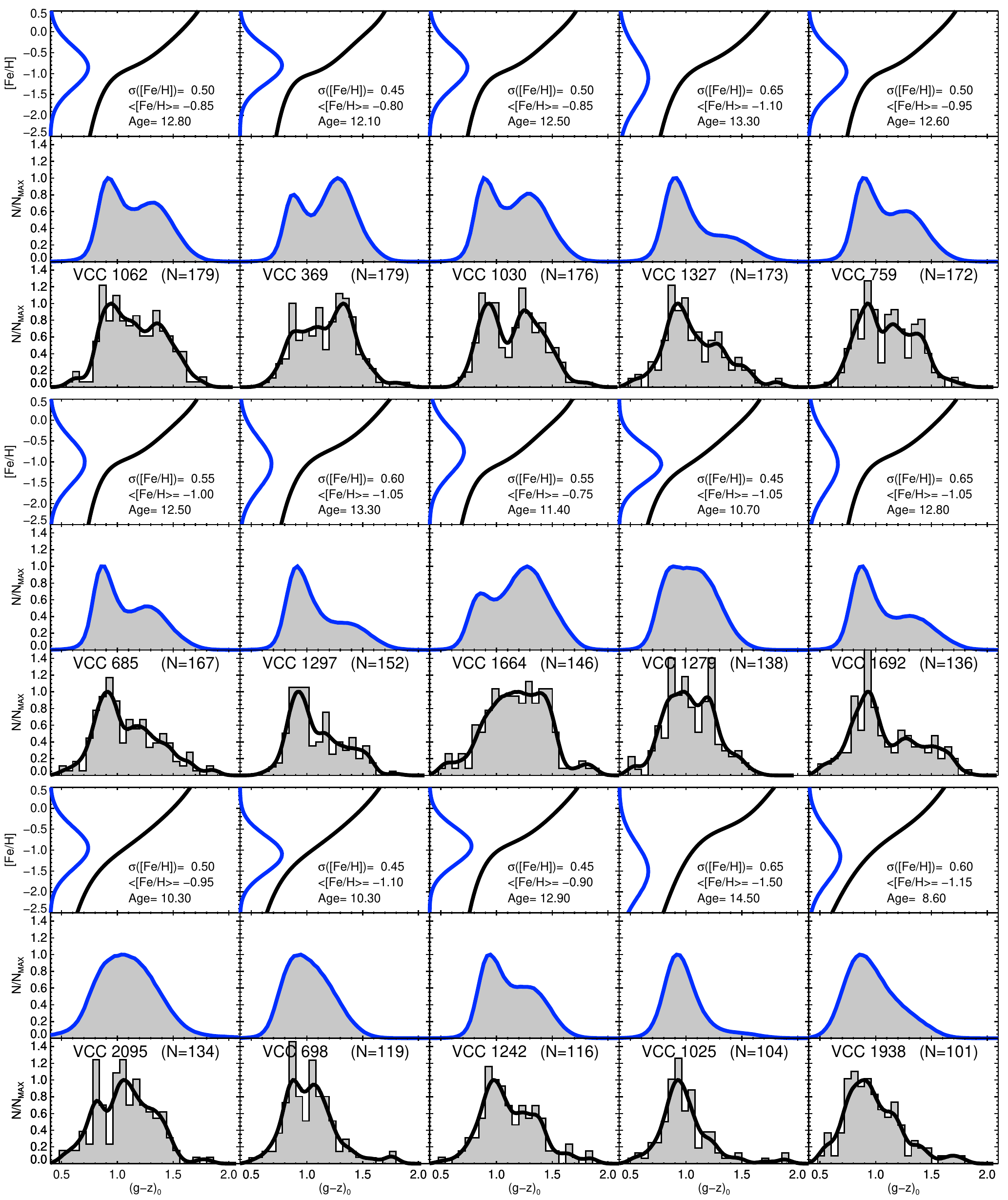}
	\caption{\emph{Continued}}
\end{figure*}
\begin{figure*}
	\figurenum{3}
	\includegraphics[width=16.0cm]{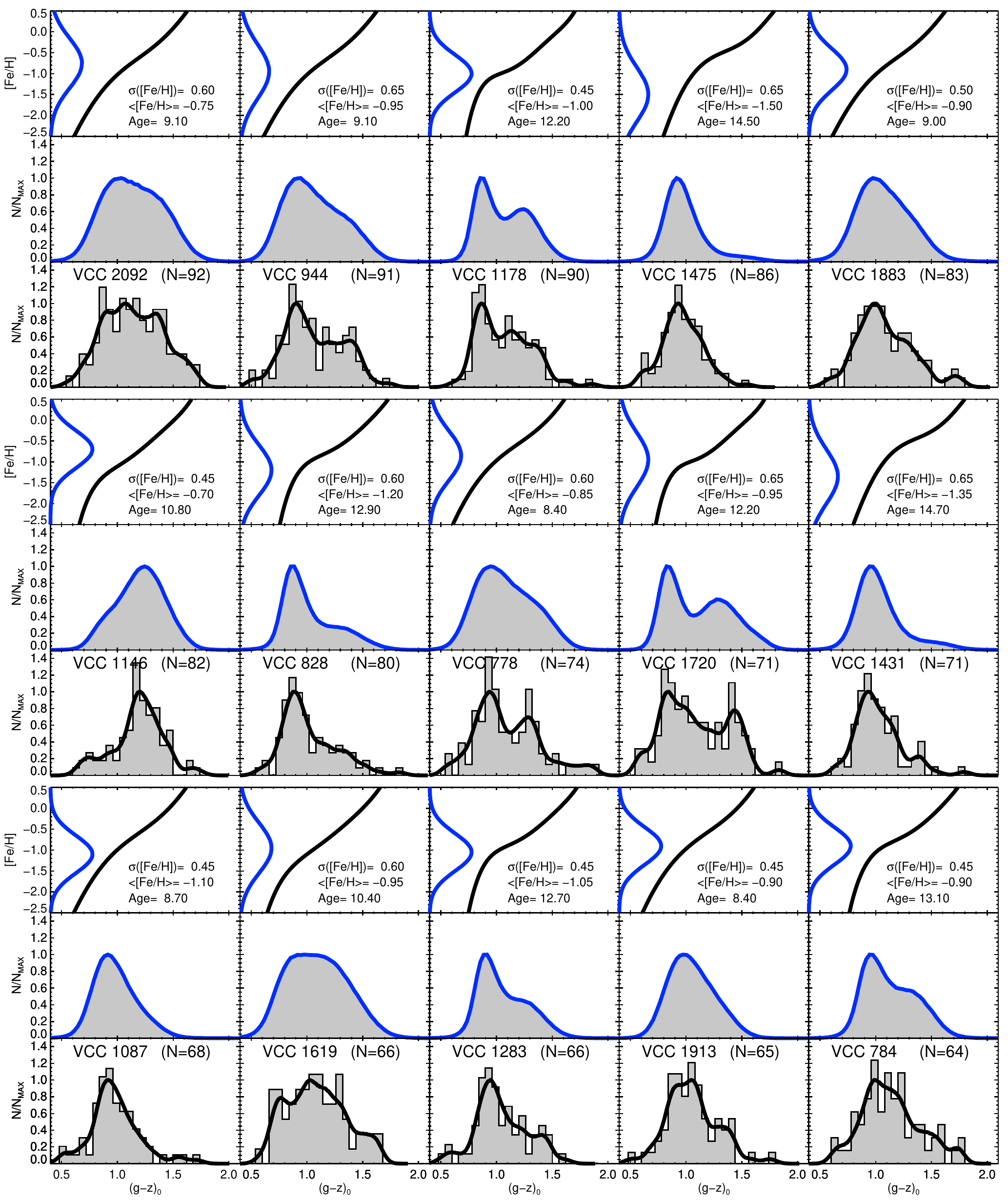}
	\caption{\emph{Continued}}
\end{figure*}
\begin{figure*}
	\figurenum{3}
	\includegraphics[width=16.0cm]{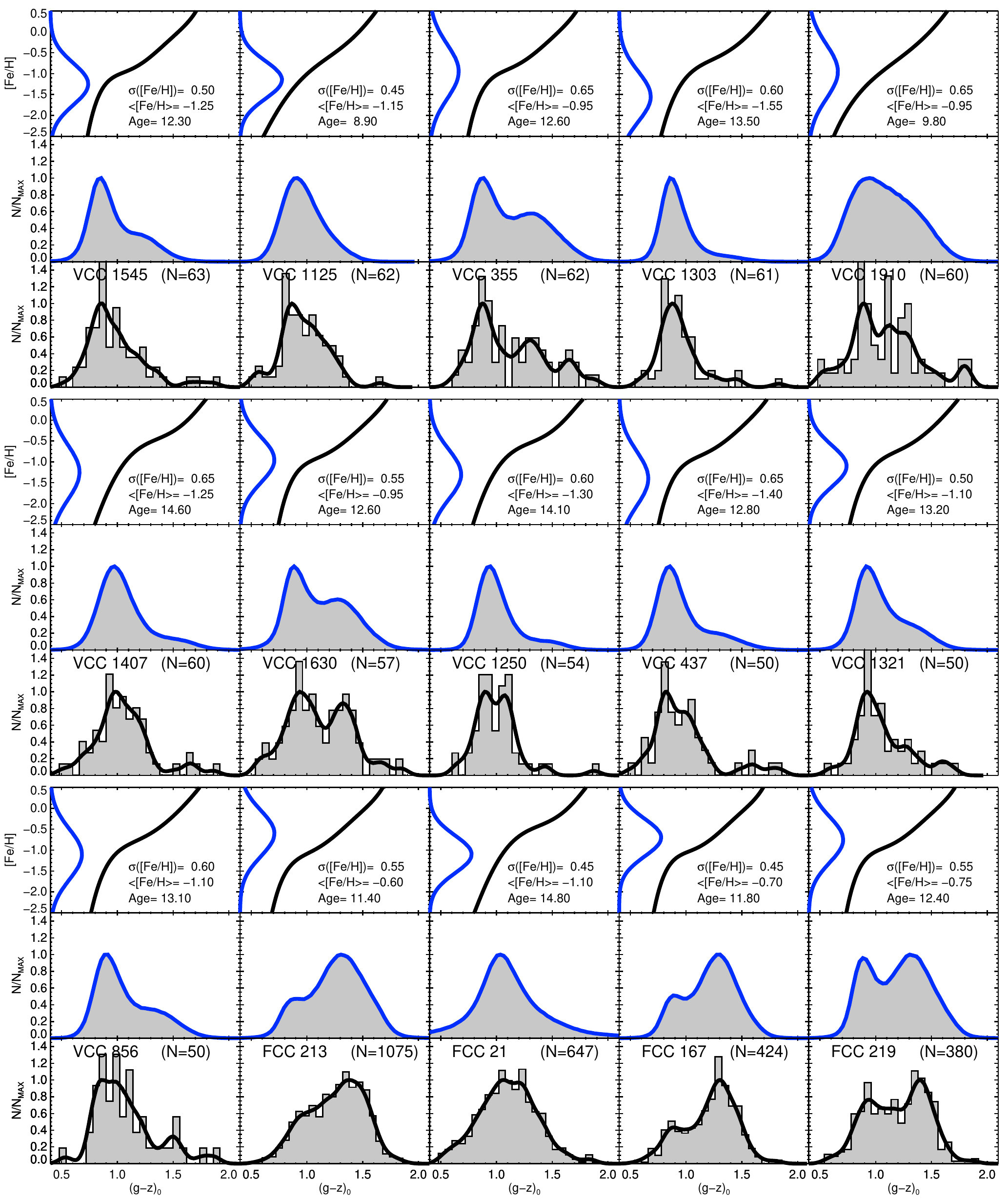}
	\caption{\emph{Continued}}
\end{figure*}
\begin{figure*}
	\figurenum{3}
	\includegraphics[width=16.0cm]{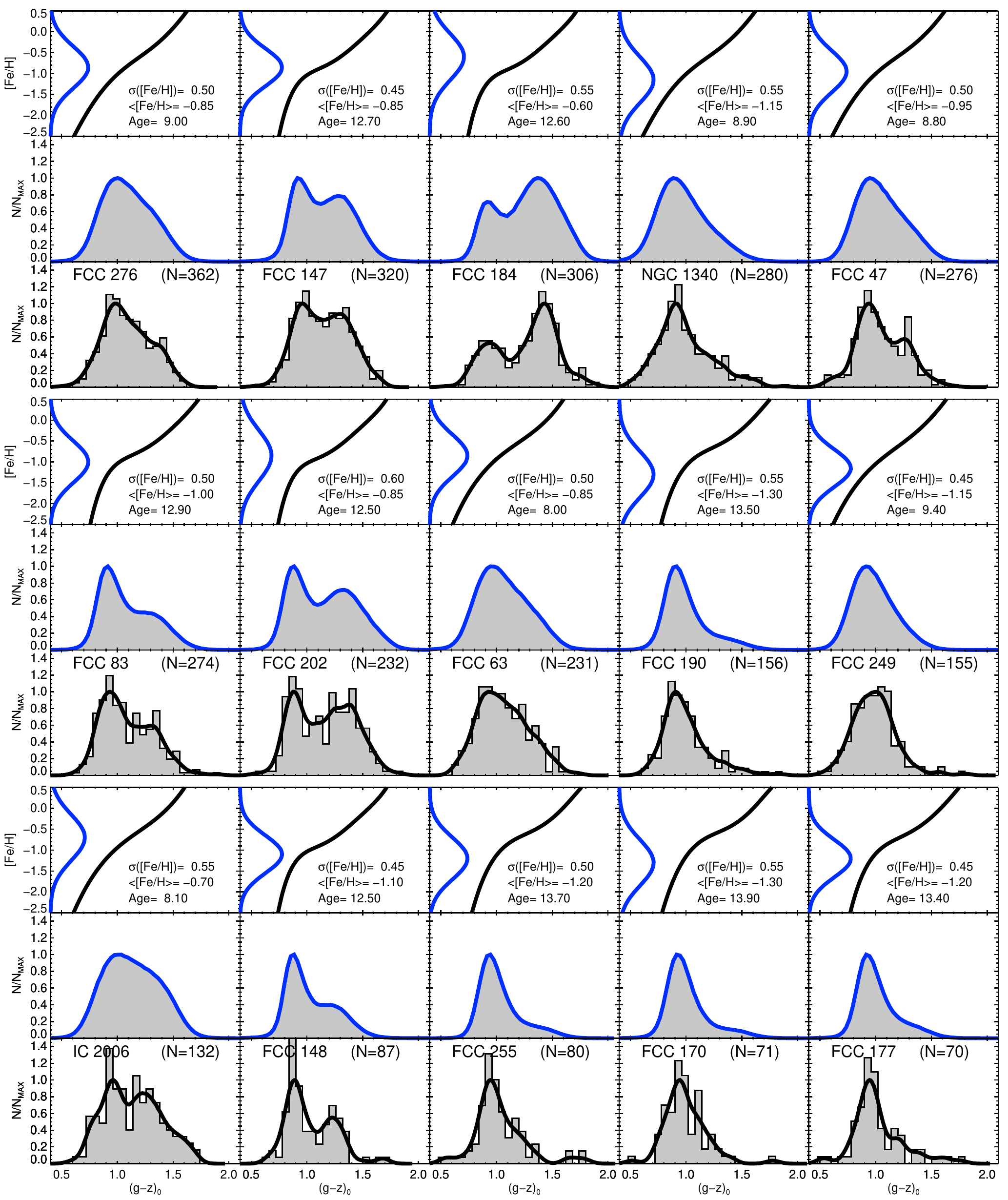}
	\caption{\emph{Continued}}
\end{figure*}
\begin{figure*}
	\figurenum{3}
	\includegraphics[width=16.0cm]{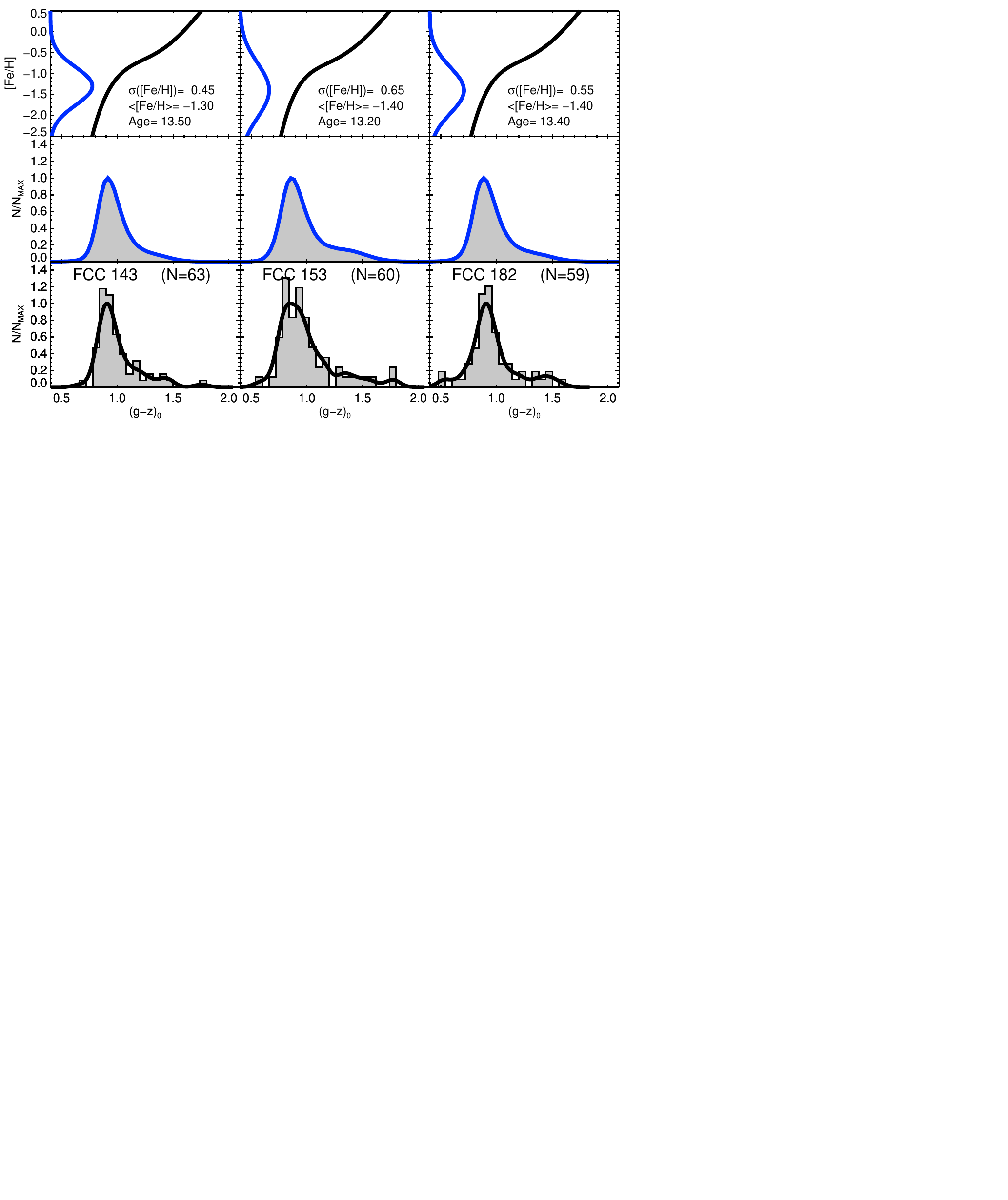}
	\caption{\emph{Continued}}
\end{figure*}

\begin{longrotatetable}
\begin{deluxetable*}{llcclccccccrrrc}
\tabletypesize{\scriptsize}
\tablecaption{Simulation Results for the ACS Virgo and Fornax Cluster Survey Galaxies\label{table2}}
\tablenum{2}
\tablewidth{0pt}
\tablehead{
\colhead{}&
\colhead{}&
\colhead{}&
\colhead{}&
\colhead{Galaxy}&
\colhead{}&
\colhead{}&
\colhead{}&
\colhead{}&
\colhead{}&
\colhead{}&
\colhead{}&
\colhead{}&
\colhead{}&
\colhead{GC}\\
\colhead{Other Name}  &
\colhead{Other name} &
\colhead{$N\textsubscript{GC}$} & 
\colhead{$M_B$} &
\colhead{Morphological} & 
\colhead{Age} & 
\colhead{${\langle{\rm}}$[Fe/H]${\rangle}$} & 
\colhead{$\sigma$([Fe/H])} & 
\colhead{$D$} & 
\colhead{$p$-value} & 
\colhead{$f\textsubscript{red}$} &
\colhead{$p(\chi^{2})$} &
\colhead{$p(DD)$} &
\colhead{$p$(kurt)} &
\colhead{CDF}\\
\colhead{}&        
\colhead{}&  
\colhead{}&  
\colhead{(mag)}&
\colhead{Type}& 
\colhead{(Gyr)}& 
\colhead{(dex)}& 
\colhead{}&
\colhead{(K-S)}&
\colhead{(K-S)}&
\colhead{(GMM)}&
\colhead{(GMM)}&
\colhead{(GMM)}&
\colhead{(GMM)}&
\colhead{Type}
}
\decimalcolnumbers
\startdata
VCC 1316&   M87, N4486& 1745&--21.6&E0              & 12.6&--0.65& 0.55&0.020&0.468& 0.57&  $<0.010$&  $<0.010$&  $<0.010$&1a\\
 VCC 731&   N4365     &  907&--21.4&E3              & 12.5&--0.75& 0.45&0.026&0.572& 0.53&  $<0.001$&  $<0.001$&  $<0.001$&1a\\
VCC 1978&   M60, N4649&  807&--21.4&S0$_1$(2)       & 12.9&--0.65& 0.55&0.026&0.667& 0.57&  $<0.001$&  $<0.001$&  $<0.001$&1a\\
VCC 1226&   M49, N4472&  765&--21.9&E2/S0$_1$(2)    & 13.0&--0.65& 0.55&0.028&0.568& 0.58&  $<0.001$&  $<0.001$&  $<0.001$&1a\\
 VCC 798&   M85, N4382&  507&--21.3&S0$_1$(3) pec   &  9.3&--0.75& 0.55&0.023&0.956& 0.31&  $<0.001$&   $0.168$&   $0.031$&1b\\
 VCC 763&   M84, N4374&  506&--21.2&E1              & 12.8&--0.90& 0.50&0.026&0.895& 0.37&  $<0.001$&  $<0.001$&   $0.003$&1a\\
VCC 1632&   M89, N4552&  456&--20.4&S0$_1$(0)       &  9.3&--0.60& 0.60&0.029&0.819& 0.51&  $<0.001$&   $0.089$&   $0.106$&1b\\
 VCC 881&   M86, N4406&  367&--21.3&S0$_1$(3)/E3    &  9.6&--0.90& 0.50&0.024&0.981& 0.14&   $0.001$&   $0.464$&   $0.835$&2b\\
VCC 1903&   M59, N4621&  308&--20.2&E4              & 11.6&--0.75& 0.50&0.033&0.893& 0.57&   $0.013$&   $0.003$&   $0.075$&1a\\
VCC 1231&   N4473     &  254&--19.9&E5              &  8.1&--0.70& 0.55&0.035&0.911& 0.38&  $<0.001$&   $0.279$&   $0.108$&1b\\
VCC 1535&   N4526     &  244&--20.6&S0$_3$(6)       & 12.4&--0.80& 0.65&0.035&0.916& 0.49&  $<0.001$&  $<0.001$&  $<0.001$&1a\\
VCC 1199&             &  228&--15.7&E2              & 13.0&--0.90& 0.60&0.026&0.998& 0.38&  $<0.001$&  $<0.001$&  $<0.001$&1a\\
VCC 1192&   N4467     &  213&--16.1&E3              & 13.2&--1.00& 0.65&0.048&0.706& 0.31&  $<0.001$&  $<0.001$&   $0.024$&1a\\
VCC 2000&   N4660     &  197&--19.1&E3/S0$_1$(3)    & 13.4&--1.15& 0.60&0.033&0.981& 0.13&  $<0.001$&   $0.366$&   $0.994$&2a\\
VCC 1154&   N4459     &  192&--19.9&S0$_3$(2)       & 12.6&--0.90& 0.45&0.028&0.998& 0.36&   $0.015$&   $0.028$&   $0.061$&1a\\
VCC 1062&   N4442     &  179&--19.6&SB0$_1$(6)      & 12.8&--0.85& 0.50&0.027&0.999& 0.42&  $<0.001$&  $<0.001$&   $0.006$&1a\\
 VCC 369&   N4267     &  179&--19.4&SB0$_1$         & 12.1&--0.80& 0.45&0.035&0.976& 0.55&   $0.002$&   $0.003$&   $0.053$&1a\\
VCC 1030&   N4435     &  176&--19.4&SB0$_1$(6)      & 12.5&--0.85& 0.50&0.040&0.936& 0.50&  $<0.001$&  $<0.001$&   $0.002$&1a\\
VCC 1327&   N4486A    &  173&--18.2&E2              & 13.3&--1.10& 0.65&0.040&0.936& 0.25&  $<0.001$&   $0.603$&   $0.564$&2a\\
 VCC 759&   N4371     &  172&--19.5&SB0$_2$(r)(3)   & 12.6&--0.95& 0.50&0.042&0.922& 0.41&   $0.001$&   $0.013$&   $0.008$&1a\\
 VCC 685&   N4350     &  167&--19.2&S0$_1$(8)       & 12.5&--1.00& 0.55&0.033&0.992& 0.27&  $<0.001$&   $0.526$&   $0.358$&2a\\
VCC 1297&   N4486B    &  152&--16.8&E1              & 13.3&--1.05& 0.60&0.040&0.966& 0.29&  $<0.001$&   $0.404$&   $0.094$&2a\\
VCC 1664&   N4564     &  146&--19.1&E6              & 11.4&--0.75& 0.55&0.037&0.988& 0.66&   $0.236$&  $<0.001$&   $0.493$&1a\\
VCC 1279&   N4478     &  138&--19.1&E2              & 10.7&--1.05& 0.45&0.043&0.958& 0.39&   $0.348$&   $0.398$&   $0.288$&1b\\
VCC 1692&   N4570     &  136&--19.4&S0$_1$(7)/E7    & 12.8&--1.05& 0.65&0.055&0.787& 0.33&  $<0.001$&  $<0.001$&   $0.016$&1a\\
VCC 2095&   N4762     &  134&--20.0&S0$_1$(9)       & 10.3&--0.95& 0.50&0.051&0.863& 0.46&   $0.251$&   $0.211$&   $0.319$&1b\\
 VCC 698&   N4352     &  119&--17.9&S0$_1$(8)       & 10.3&--1.10& 0.45&0.049&0.936& 0.06&  $<0.001$&  $<0.001$&   $0.995$&2b\\
VCC 1242&   N4474     &  116&--18.5&S0$_1$(8)       & 12.9&--0.90& 0.45&0.046&0.962& 0.14&   $0.008$&   $0.673$&   $0.759$&2a\\
VCC 1025&   N4434     &  104&--18.8&E0/S0$_1$(0)    & 14.5&--1.50& 0.65&0.045&0.983& 0.04&  $<0.001$&   $0.542$&   $1.000$&2a\\
VCC 1938&   N4638     &  101&--19.2&S0$_1$(7)       &  8.6&--1.15& 0.60&0.032&1.000& 0.11&   $0.003$&   $0.517$&   $0.934$&2b\\
VCC 2092&   N4754     &   92&--19.7&SB0$_1$(5)      &  9.1&--0.75& 0.60&0.035&1.000& 0.43&   $0.108$&   $0.148$&   $0.035$&1b\\
 VCC 944&   N4417     &   91&--19.0&S0$_1$(7)       &  9.1&--0.95& 0.65&0.049&0.977& 0.38&   $0.009$&   $0.010$&   $0.060$&1b\\
VCC 1178&   N4464     &   90&--17.7&E3              & 12.2&--1.00& 0.45&0.048&0.982& 0.31&  $<0.001$&   $0.422$&   $0.657$&1a\\
VCC 1475&   N4515     &   86&--17.9&E2              & 14.5&--1.50& 0.65&0.054&0.956& 0.05&   $0.186$&  $<0.001$&   $0.884$&2a\\
VCC 1883&   N4612     &   83&--18.6&RSB0$_{1/2}$    &  9.0&--0.90& 0.50&0.043&0.998& 0.14&   $0.052$&   $0.646$&   $0.768$&2b\\
VCC 1146&   N4458     &   82&--18.2&E1              & 10.8&--0.70& 0.45&0.085&0.577& 0.82&   $0.032$&   $0.748$&   $0.754$&1b\\
 VCC 828&   N4387     &   80&--18.6&E5              & 12.9&--1.20& 0.60&0.046&0.995& 0.20&  $<0.001$&   $0.689$&   $0.911$&2a\\
 VCC 778&   N4377     &   74&--18.7&S0$_1$(3)       &  8.4&--0.85& 0.60&0.058&0.962& 0.14&   $0.022$&   $0.697$&   $0.688$&2b\\
VCC 1720&   N4578     &   71&--18.9&S0$_{1/2}$(4)   & 12.2&--0.95& 0.65&0.060&0.952& 0.40&   $0.004$&   $0.011$&   $0.015$&1a\\
VCC 1431&   I3470     &   71&--16.7&dE0,N           & 14.7&--1.35& 0.65&0.044&0.999& 0.09&   $0.004$&   $0.658$&   $0.993$&2a\\
VCC 1087&   I3381     &   68&--16.9&dE3,N           &  8.7&--1.10& 0.45&0.061&0.959& 0.07&   $0.009$&  $<0.001$&   $0.988$&2b\\
VCC 1619&   N4550     &   66&--18.6&E7/S0$_1$(7)    & 10.4&--0.95& 0.60&0.052&0.992& 0.37&   $0.054$&   $0.264$&   $0.058$&1b\\
VCC 1283&   N4479     &   66&--17.9&SB0$_2$(2)      & 12.7&--1.05& 0.45&0.051&0.994& 0.28&   $0.082$&   $0.026$&   $0.459$&2a\\
VCC 1913&   N4623     &   65&--18.1&E7              &  8.4&--0.90& 0.45&0.051&0.995& 0.14&   $0.151$&   $0.759$&   $0.870$&2b\\
 VCC 784&   N4379     &   64&--18.4&S0$_1$(2)       & 13.1&--0.90& 0.45&0.055&0.989& 0.24&   $0.086$&   $0.028$&   $0.405$&2a\\
VCC 1545&   I3509     &   63&--16.3&E4              & 12.3&--1.25& 0.50&0.047&0.999& 0.06&   $0.001$&   $0.665$&   $0.993$&2a\\
VCC 1125&   N4452     &   62&--17.9&S0$_1$(9)       &  8.9&--1.15& 0.45&0.057&0.985& 0.06&   $0.077$&  $<0.001$&   $0.950$&2b\\
 VCC 355&   N4262     &   62&--18.7&SB0$_{2/3}$     & 12.6&--0.95& 0.65&0.072&0.890& 0.34&  $<0.001$&   $0.205$&   $0.095$&1a\\
VCC 1303&   N4483     &   61&--18.1&SB0$_1$(5)      & 13.5&--1.55& 0.60&0.056&0.990& 0.07&  $<0.001$&   $0.687$&   $0.999$&2a\\
VCC 1910&   I809      &   60&--17.0&dE1,N           &  9.8&--0.95& 0.65&0.062&0.972& 0.16&   $0.007$&  $<0.001$&   $0.621$&2b\\
VCC 1407&   I3461     &   60&--15.8&dE2,N           & 14.6&--1.25& 0.65&0.061&0.974& 0.09&   $0.002$&  $<0.001$&   $0.983$&2a\\
VCC 1630&   N4551     &   57&--18.3&E2              & 12.6&--0.95& 0.55&0.047&0.999& 0.40&   $0.271$&   $0.464$&   $0.374$&1a\\
VCC 1250&   N4476     &   54&--18.5&S0$_3$(5)       & 14.1&--1.30& 0.60&0.091&0.749& 0.02&  $<0.001$&   $0.459$&   $1.000$&2a\\
 VCC 437&   U7399A    &   50&--16.8&dE5,N           & 12.8&--1.40& 0.65&0.065&0.982& 0.10&  $<0.001$&  $<0.001$&   $0.995$&2a\\
VCC 1321&   N4489     &   50&--18.2&S0$_1$(1)       & 13.2&--1.10& 0.50&0.058&0.995& 0.19&   $0.004$&   $0.312$&   $0.794$&2a\\
 VCC 856&   I3328     &   50&--17.0&dE1,N           & 13.1&--1.10& 0.60&0.056&0.997& 0.17&   $0.002$&   $0.084$&   $0.890$&2a\\
 FCC 213&  N1399   & 1075&--21.0&E0              & 11.4&--0.60& 0.55&0.027&0.399& 0.64&  $<0.010$&  $<0.010$&  $<0.010$&1a\\
  FCC 21&  N1316   &  647&--22.3&S0$_3$(pec)     & 14.8&--1.10& 0.45&0.068&0.005& 0.12&   $0.242$&   $0.560$&   $0.416$&2a\\
 FCC 167&  N1380   &  424&--20.4&S0/a            & 11.8&--0.70& 0.45&0.039&0.540& 0.71&  $<0.001$&  $<0.001$&   $0.052$&1a\\
 FCC 219&  N1404   &  380&--20.7&E2              & 12.4&--0.75& 0.55&0.043&0.476& 0.55&  $<0.001$&  $<0.001$&  $<0.001$&1a\\
 FCC 276&  N1427   &  362&--19.7&E4              &  9.0&--0.85& 0.50&0.034&0.802& 0.35&  $<0.001$&  $<0.001$&   $0.003$&1b\\
 FCC 147&  N1374   &  320&--19.6&E0              & 12.7&--0.85& 0.45&0.030&0.927& 0.46&  $<0.001$&  $<0.001$&  $<0.001$&1a\\
 FCC 184&  N1387   &  306&--19.2&SB0             & 12.6&--0.60& 0.55&0.055&0.303& 0.65&  $<0.001$&  $<0.001$&  $<0.001$&1a\\
NGC 1340&  N1344   &  280&--20.4&E5              &  8.9&--1.15& 0.55&0.038&0.797& 0.17&  $<0.001$&   $0.086$&   $0.940$&2b\\
  FCC 47&  N1336   &  276&--18.1&E4              &  8.8&--0.95& 0.50&0.042&0.712& 0.33&  $<0.001$&  $<0.001$&   $0.228$&1b\\
  FCC 83&  N1351   &  274&--19.2&E5              & 12.9&--1.00& 0.50&0.036&0.868& 0.34&  $<0.001$&   $0.022$&   $0.203$&1a\\
 FCC 202&  N1396   &  232&--16.3&d:E6,N          & 12.5&--0.85& 0.60&0.035&0.937& 0.51&  $<0.001$&   $0.003$&  $<0.001$&1a\\
  FCC 63&  N1339   &  231&--18.8&E4              &  8.0&--0.85& 0.50&0.025&0.999& 0.30&  $<0.001$&   $0.179$&   $0.129$&1b\\
 FCC 190&  N1380B  &  156&--18.1&SB0             & 13.5&--1.30& 0.55&0.030&0.999& 0.09&  $<0.001$&   $0.448$&   $1.000$&2a\\
 FCC 249&  N1419   &  155&--18.2&E0              &  9.4&--1.15& 0.45&0.053&0.775& 0.01&   $0.002$&  $<0.001$&   $0.999$&2b\\
 IC 2006&  ESO 359-G07&  132&--19.4&E               &  8.1&--0.70& 0.55&0.037&0.993& 0.42&   $0.004$&   $0.065$&   $0.019$&1b\\
 FCC 148&  N1375   &   87&--18.0&S0(cross)       & 12.5&--1.10& 0.45&0.058&0.925& 0.34&  $<0.001$&   $0.608$&   $0.778$&1a\\
 FCC 255&  ESO 358-G50&   80&--17.8&S0$_1$(6),N     & 13.7&--1.20& 0.50&0.045&0.996& 0.05&  $<0.001$&  $<0.001$&   $0.999$&2a\\
 FCC 170&  N1381   &   71&--18.8&S0(9)(boxy)     & 13.9&--1.30& 0.55&0.054&0.983& 0.02&   $0.004$&   $0.775$&   $1.000$&2a\\
 FCC 177&  N1380A  &   70&--18.4&S0$_2$(9)(cross)& 13.4&--1.20& 0.45&0.056&0.976& 0.11&  $<0.001$&   $0.163$&   $0.987$&2a\\
 FCC 143&  N1373   &   63&--17.2&E3              & 13.5&--1.30& 0.45&0.081&0.789& 0.10&  $<0.001$&   $0.697$&   $1.000$&2a\\
 FCC 153&  ESO 358-G26&   60&--18.6&S0$_1$(9)       & 13.2&--1.40& 0.65&0.051&0.997& 0.13&  $<0.001$&   $0.448$&   $0.993$&2a\\
 FCC 182&             &   59&--16.6&SB0 pec         & 13.4&--1.40& 0.55&0.064&0.964& 0.14&  $<0.001$&  $<0.001$&   $0.964$&2a\\
\enddata
\tablecomments{(1) Galaxy VCC and FCC number. (2) Other name. (3) The total number of observed GCs.  (4) Galaxy $B$-band magnitude~\citep{2011ApJ...726...31G}. (5) Morphological classification obtained from \cite{2004ApJS..153..223C} and \cite{2007ApJS..169..213J}. (6) The best-fit age. (7) The best-fit mean [Fe/H]. (8) The best-fit dispersion of MDF. (9) K-S statistic \textit{D}. (10) K-S probability. (11) Red GC fraction in $g-z$ color distributions from GMM analysis. (12)--(14) $p$-values from GMM analysis. $p(\chi^{2}$), $p(DD)$, and $p$(kurt) are based on the likelihood-ratio test, separation of the peaks, and kurtosis, respectively. The lower $p$-values suggest more significant bimodality. (15) GC CDF type.}
\end{deluxetable*}
\end{longrotatetable}

\begin{deluxetable*}{lccrccc}
\tabletypesize{\normalsize}
\tablewidth{0pt}
\tablenum{3}
\tablecaption{Classification of Color Distribution Function Types and Mean Values of the Simulated Parameters for Each Type\label{table:table3}}
\tablehead{
\colhead{Type} &
\colhead{Selection Criteria} &
\colhead{$N\textsubscript{Gal}$} &
\colhead{Age} &
\colhead{${\langle{\rm}}$[Fe/H]$\rangle$} &
\colhead{$\sigma$([Fe/H])} &
\colhead{N(S0)/N(E)}\\
\colhead{}&
\colhead{}&
\colhead{}& 
\colhead{(Gyr)}& 
\colhead{(dex)}& 
\colhead{}&
\colhead{}
}
\decimalcolnumbers
\startdata
Type 1a&$t\textsubscript{best-fit}>11$ Gyr, $f_{\rm red}>0.3$       &28& 12.5$\pm$0.09&--0.84$\pm$0.026& 0.54$\pm$0.013& 1.00\\
Type 1b&$t\textsubscript{best-fit}\leq11$ Gyr, $f_{\rm red}>0.3$    &13&  9.3$\pm$0.27&--0.83$\pm$0.038& 0.54$\pm$0.017& 0.86\\
Type 2a&$t\textsubscript{best-fit}>11$ Gyr, $f_{\rm red}\leq0.3$    &26& 13.5$\pm$0.13&--1.22$\pm$0.035& 0.56$\pm$0.015& 1.60\\
Type 2b&$t\textsubscript{best-fit}\leq11$ Gyr, $f_{\rm red}\leq0.3$ &11&  9.1$\pm$0.18&--1.03$\pm$0.038& 0.51$\pm$0.022& 1.20\\
\enddata
\tablecomments{(1) GC CDF type. (2) Selection criteria. $f\textsubscript{red}$ is the red GC fraction in the $g-z$ color distributions. In the GMM model, the homoscedastic case is adopted to derive $f\textsubscript{red}$. (3) The number of galaxies belonging to each type.  (4) The mean value of the best-fit ages. (5) The mean value of the best-fit ${\langle{\rm}}$[Fe/H]${\rangle}$. (6) The mean value of the best-fit MDF dispersions. (7) The number ratio of lenticular galaxies to elliptical galaxies}
\end{deluxetable*}

While Types 1a and 2a have the characteristics of the old CDF models (i.e., clear bimodality of Type 1a and a sharp blue peak of Type 2a), Types 1b and 2b do not have such distinct morphological features.
Thus, Types 1b and 2b have a relatively younger best-fit age ($8\sim11$ Gyr), and the age variation due to the different choice of $\sigma$({[Fe\slash H]}) is larger than that of Types 1a and 2a.
In other words, there is quite a bit of degeneracy between three parameters (age, ${\langle{\rm}}${[Fe\slash H]}${\rangle}$, and $\sigma$({[Fe\slash H]})) for Types 1b and 2b.
Even with similar $N\textsubscript{GC}$, the \emph{p}-value contours are less compact compared to Types 1a and 2a, indicating that the assumption of old and coeval GCs may not hold for Types 1b and 2b.
We attempt to address the nature of Types 1b and 2b in Section 5.

\begin{figure*}
\begin{center}
\includegraphics[width=12.3cm]{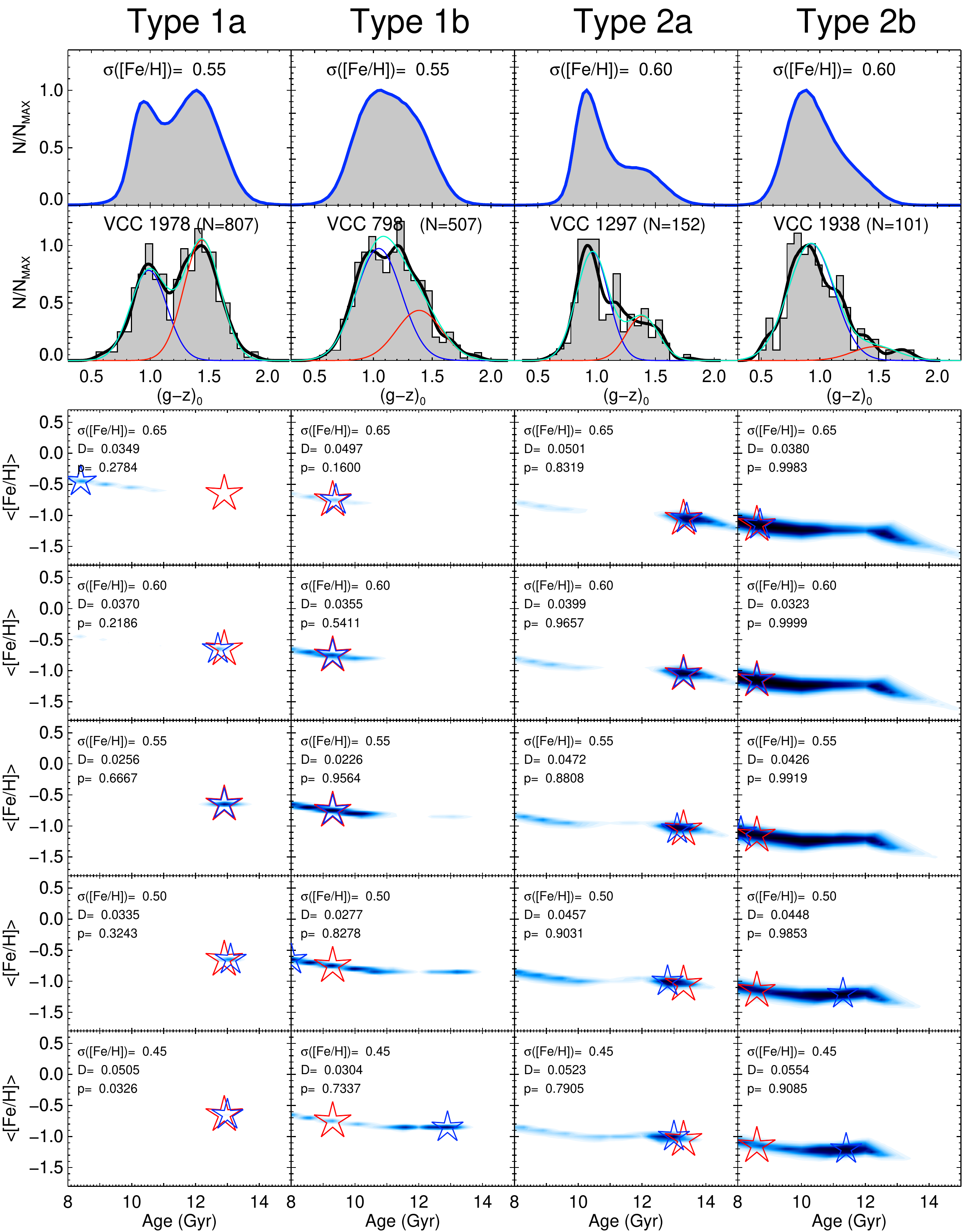}
\caption{First two rows: the four types of GC CDF morphologies. The simulated $g-z$ distributions are compared with the representative observations. (First row): the blue histograms are the $g-z$ CDFs of our best models. The metallicity dispersion of the best model is denoted. (Second row): the observed $g-z$ CDFs for four types of CDFs. The blue and red lines are the Gaussian fits from the GMM test for the homoscedastic case. Cyan lines represent the sums of the blue and red lines. The Gaussian kernel density estimations (black solid lines) are the same as in Figure 2. Bottom five rows: the effect of differing $\sigma$([Fe/H]). The confidence contours for age and ${\langle{\rm}}$[Fe/H]${\rangle}$, based on the K-S test between models and observations. The blue stars show the best-fit age and best-fit ${\langle{\rm}}$[Fe/H]${\rangle}$ for each case of $\sigma$([Fe/H]). The red stars represent the best-fit age and ${\langle{\rm}}$[Fe/H]${\rangle}$, which are finally selected using $\sigma$([Fe/H]) as a free parameter. The contour levels are the same as in Figure 1.}
\label{fig4}
\end{center}
\end{figure*}

\section{T\MakeLowercase{he} D\MakeLowercase{erived} P\MakeLowercase{arameters} \MakeLowercase{as} F\MakeLowercase{unctions} \MakeLowercase{of} H\MakeLowercase{ost} G\MakeLowercase{alaxy} P\MakeLowercase{roperties}}
\label{thederivedparametersasfunctionsofhostgalaxyproperties}

\subsection{Best-fit Parameters and Galaxy Luminosity}
\label{best-fitparametersandgalaxyluminosity}

\begin{figure}
\begin{center}
\includegraphics[width=8.5cm]{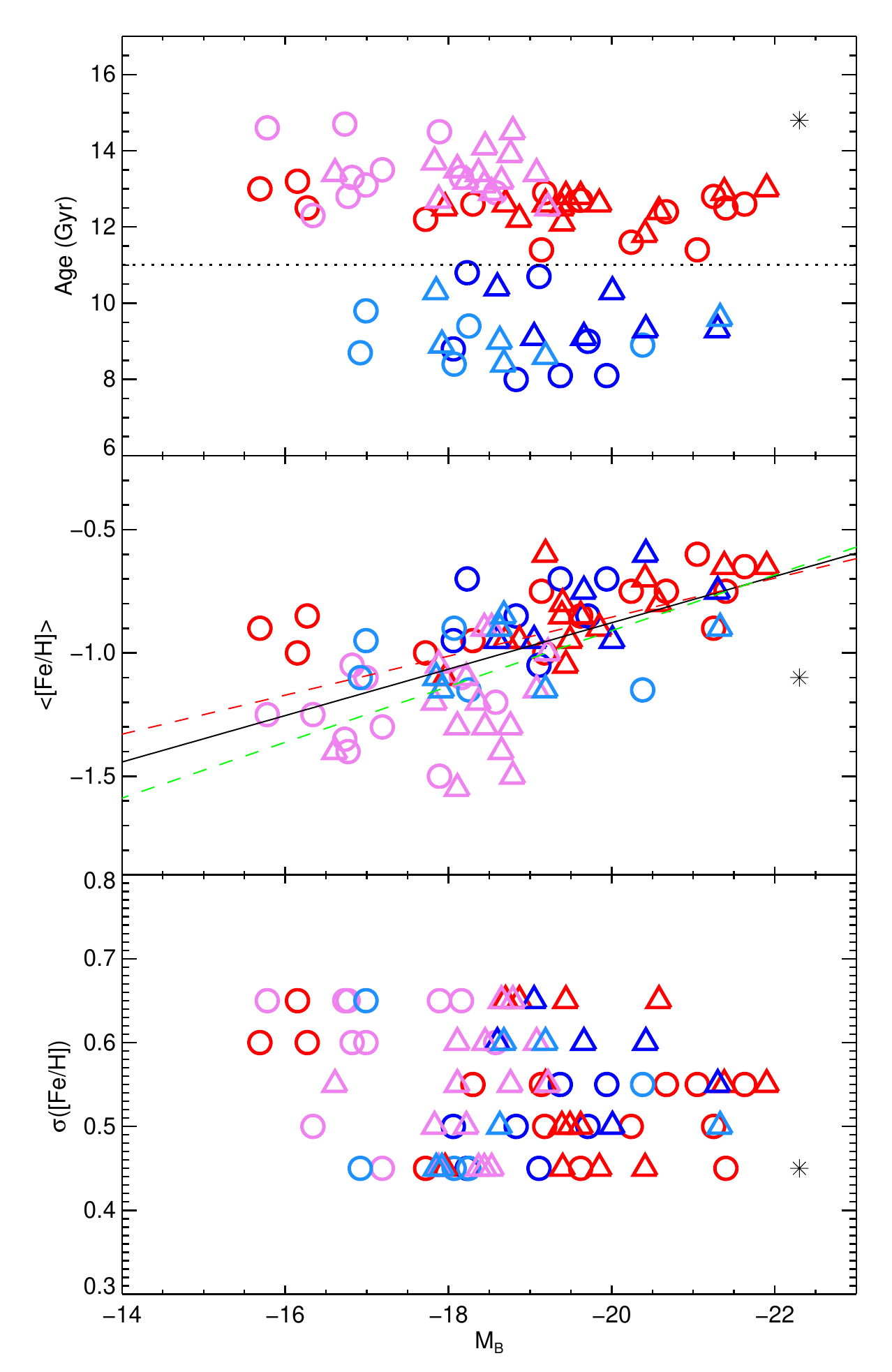}   
\caption{Best-fit age (top), best-fit ${\langle{\rm}}$[Fe/H]${\rangle}$ (middle), and best-fit $\sigma$([Fe/H]) (bottom) as a function of the host galaxy luminosity. 
The circles and triangles represent elliptical and lenticular galaxies, respectively.
In the top panel, the horizontal dotted line at 11 Gyr defines the demarcation between old and old\,+\,young GC systems.
The red (Type 1a) and violet (Type 2a) symbols represent the old GC systems ($>11$ Gyr), while the blue (Type 1b) and light blue (Type 2b) ones denote the old\,+\,young GC systems ($<11$ Gyr).
In the middle panel, the linear fit to all the galaxies is marked as a black solid line. 
The red and green dashed lines represent the linear fit to elliptical and lenticular galaxies, respectively.
E/S0 and S0/E type galaxies in Table 2 are dealt with as lenticular galaxies.
FCC 21 with $p_{(K-S)}<0.05$ is overplotted as an asterisk and excluded from the fit.}
\label{fig5}
\end{center}
\end{figure}

Figure~\ref{fig5} shows the best-fit age, ${\langle{\rm}}${[Fe\slash H]}${\rangle}$, and $\sigma$({[Fe\slash H]}) of the entire GC systems as functions of the host galaxy luminosity.
The best-fit age does not show a clear correlation with the host luminosity.
The distribution of the best-fit age can be broken into two groups at 11 Gyr, above which the $g-z$ CMRs are highly inflected (see Figure 1).
A detailed explanation for the galaxies with a model-derived age less than 11 Gyr is given in Section 5.
Unlike the best-fit age, the best-fit ${\langle{\rm}}${[Fe\slash H]}${\rangle}$ shows a tight correlation with the host galaxy luminosity.
The derived $M_{B}$--${\langle{\rm}}${[Fe\slash H]}${\rangle}$ slope for the whole galaxy sample is $-0.094 \pm 0.015$, while the slope excluding the galaxies with $t\textsubscript{best-fit}<11$ Gyr is $-0.108 \pm 0.019$.
The best-fit $\sigma$({[Fe\slash H]}) distribution has no obvious correlation with the host luminosity.

We note that the ACSVCS and ACSFCS data have a limitation in radial coverage for large galaxies due to the small field of view of ACS\slash WFC.
For instance, ACS\slash WFC covers only the inner region ($\sim$\,1\,$R_{\rm eff}$) of VCC 1316 (M87), the largest galaxy in our sample.
The mean metallicity of GCs is known to be higher in the inner region~\citep[e.g.,][]{1996AJ....111.1529G,1999ApJ...513..733K,2001AJ....121.1992F,2009ApJ...703..939H,2012MNRAS.420...37B,2018MNRAS.479.4760F}.
Thus, for some very large galaxies the mean metallicity of observed GCs is determined to be higher than their mean metallicity for all the GCs.
However, most of our sample galaxies are fairly covered by the ACS\slash WFC field of view~\citep[see Figure 2 in][]{2013ApJ...769..145W} and that limitation does not have a significant effect on the overall trend of the $M_{B}$--${\langle{\rm}}${[Fe\slash H]}${\rangle}$ relation.
Moreover, observations of the color bimodality of GC systems in early-type galaxies reveal that the red GC fraction gets higher with host galaxy luminosity. The faintest galaxies in ACSVCS on average have a 15\,$\%$ fraction of red GCs, and the fraction increases to 60\,$\%$ for the brightest galaxies~\citep{2006ApJ...639...95P}.
Hence, the main driver behind the $M_{B}$--${\langle{\rm}}${[Fe\slash H]}${\rangle}$ relation is most likely this changing fraction of red GCs rather than the field of view bias of ACS\slash WFC.

\subsection{Best-fit Parameters and Galaxy Morphological Type}
\label{best-fitparametersandgalaxymorphologicaltype}

In the top panel of Figure 5, the best-fit age derived from our GC color distribution model shows no difference depending on the morphological types (E vs. S0) of the host galaxies. About 30\,$\%$ among both E and S0 galaxies are identified as $t\textsubscript{best-fit}<11$ Gyr; 33\,$\%$ for E galaxies and 29\,$\%$ for S0 galaxies.
In the middle panel, the best-fit ${\langle{\rm}}${[Fe\slash H]}${\rangle}$ of both E and S0 galaxies increases as the host galaxy luminosity increases.
The slope of the relation for the E galaxies ($-0.079\pm0.018$) is slightly shallower than that of the S0 galaxies ($-0.113\pm0.027$).
In the bottom panel, the mean $\sigma$({[Fe\slash H]}) of both E and S0 galaxies shows almost the same value ($\sim$$0.54$).

\citet{2011A&A...525A..20C} estimated the ages of the GC systems in 14 E\slash S0 galaxies from the ($g-k$) vs. ($g-z$) diagram, and found that their S0 galaxies, compared to the E galaxies, have preferentially younger blue GCs. The discrepancy in the GC age vs. galaxy morphology relations between Chies-Santos et al. and ours seems to be due to the different choices of age estimation method. 
Moreover, the galaxy morphological classification depends fairly on references even for such well-studied nearby galaxies. For instance, VCC 1978 (M60), a galaxy without young GCs in both studies, is classified as E in the Third Reference Catalogue of Bright Galaxies (RC3\footnote{\scriptsize\url{https://heasarc.nasa.gov/W3Browse/all/rc3.html}})~\citep{1991RC3.9.C...0000d} but as S0 in the Revised Shapley-Ames Catalog of Bright Galaxies (RSA\footnote{\scriptsize\url{https://ned.ipac.caltech.edu/level5/Shapley_Ames/frames.html}})~\citep{1981rsac.book.....S}.
VCC 2000 (NGC 4660), a galaxy with young GCs in Chies-Santos et al. but old GCs only in ours, is classified as E in the RC3 and RSA as well as~\citet{2007MNRAS.379..401E} and~\citet{2007MNRAS.379..418C}, but as S0 in~\citet{2009ApJS..182..216K}.

\section{GC S\MakeLowercase{ystems with} $\MakeLowercase{t}\textsubscript{\MakeLowercase{best-fit}}<11$ G\MakeLowercase{yr}}
\label{globularclustersystemswitht_best-fit11gyr}

\subsection{Dilution of Bimodality}
\label{dilutionofbimodality}

In Figure 5, the GC systems can be divided into two groups; old ($11\sim15$ Gyr) and relatively younger ($<11$ Gyr) systems.
The best-fit ages of the GC systems in 24 galaxies (out of 78 galaxies) are estimated to be younger than 11 Gyr.
The model fails to reproduce FCC 21's observed CDF.
Thus, a total of 25 galaxies ($\sim$\,30\,$\%$) among our sample do not fit into our old ($>11$ Gyr) and coeval assumption.
There are two ways to interpret these GC systems.
First, one can simply accept the result as it is: that GCs in each galaxy are younger than 11 Gyr, with coeval formation epochs.
Indeed, the match between the younger model CDFs and the observed CDFs is fairly good (\emph{p}-value $>0.05$).
But there is another possibility: the GC systems have newly formed GC populations along with the underlying old GCs.
There are many galaxies possessing young or intermediate-age GC populations suggested by spectroscopy~\citep[e.g.,][]{2001MNRAS.322..643G,2003ApJ...585..767L,2003AJ....125..626S,2008MNRAS.386.1443B,2010ApJ...708.1335W,2018ApJ...859..108K,2018MNRAS.479..478S} and UV or IR photometry~\citep[e.g.,][]{2002A&A...391..453P,2006AJ....131..866S,2007ApJ...661..768H,2011A&A...525A..20C,2012MNRAS.420.1317G,2014ApJ...790..122T}. 
Even with young or intermediate-age GCs, however, the galaxies are dominated by old ($\gtrsim10$ Gyr) GCs.
While our old systems (Types 1a and 2a) show typical CDF shapes, the younger systems (Types 1b and 2b) show broad unimodal shapes in their CDFs (see Figures 3 and 4).
We suspect that such ``diluted'' bimodality is the result of contamination of underlying, old CDFs (i.e., Types 1a and 2a) by young or intermediate-age GCs.

\subsection{Notes on Galaxies Presumably Containing Younger GCs}
\label{notesongalaxiespresumablycontainingyoungerglobularclusters}

One of the most common sources of new stellar populations is mergers.
Peculiar morphological structures (e.g., tidal tails, shells, kinematically decoupled cores, and isophotal twist) can be interpreted as a remnant of recent mergers~\citep[e.g.,][]{1972ApJ...178..623T,1991ARA&A..29..239D,1992ApJ...393..484B,1998A&A...332...33M,2003ApJ...597..893N,2004ApJ...614L..29L,2004MNRAS.354..522M,2008AJ....135.2406S,2009ApJ...705..920H,2010ApJ...723..818H,2012ApJ...749L...1K,2012MNRAS.421.3612T}.
In the following, we briefly discuss the observational work in the literature on galaxies with $t\textsubscript{best-fit}<11$ Gyr, focusing on possible vestiges of their recent mergers and/or star formation.

VCC 798 (NGC 4382). The morphological feature of this lenticular galaxy shows that it experienced a recent merger~\citep{2018ApJ...859..108K}.
This galaxy shows possible dust patches and likely hosts a stellar disk~\citep{2006ApJS..164..334F}.
\citet{2002MNRAS.330..547T} estimated the age of VCC 798 based on the H\emph{$\beta$} and {[MgFe]} indices.
Its luminosity-weighted age, 1.6 Gyr, points to recent star formation.
\citet{2018ApJ...859..108K} estimated the spectroscopic ages of 20 GCs with the Gemini Multi-Object Spectrograph.
Among their sample GCs, 11 GCs are classified as an intermediate-age population ($\sim$$3.7$ Gyr).

VCC 1632 (NGC 4552). This lenticular galaxy is well known for its active galactic nucleus~\citep{1999ApJ...519..117C} and nuclear outflow activity~\citep{2006ApJ...648..947M}. \citet{2006ApJ...644..155M} found a gas-stripping feature in the galaxy's interstellar medium with a \emph{Chandra} observation.

VCC 881 (NGC 4406). This lenticular galaxy has a dust trail in the central region that is considered to be due to ram-pressure-stripping by its companion VCC 882~\citep{2000AJ....120..733E}. The H\emph{$\alpha$} feature produced by a collision with NGC 4438~\citep{2008ApJ...687L..69K} and a ram-pressure-stripped tail against the Virgo intracluster medium~\citep{2008ApJ...688..208R} are also reported.
\citet{2012ApJ...757..184P} estimated the spectroscopic age of 8 GCs with the Faint Object Camera and Spectrograph on the Subaru telescope.
The estimated mean age of eight GCs is reported to be 9.7 Gyr.

VCC 1231 (NGC 4473). This elliptical galaxy is well known for its kinematically distinct components, namely double peaks in the velocity dispersion map.
The galaxy shows peculiarities in both its kinematical and photometrical data, which could originate from a recent merger~\citep{2003ApJ...596..903P}.
The spectroscopic result by \citet{2011MNRAS.417.1643K} shows that the central part of this galaxy consists of younger and more metal-rich stellar populations compared to the outer part.
\citet{2015MNRAS.452.2208A} suggested that GCs follow peculiar stellar kinematics, which is evidence of a gas-rich major merger event.

VCC 1279 (NGC 4478). This elliptical galaxy contains a kinematically decoupled core~\citep{2001MNRAS.326..473H,2007MNRAS.379..401E} and a nuclear stellar disk~\citep{2004MNRAS.354..753M,2010A&A...518A..32M,2010MNRAS.407..969L}. The nuclear disk is younger ($\sim$\,6 Gyr), more metal-rich ({[Z\slash H]}\,$\sim$\,0.4), and less $\alpha$-enhanced ({[$\alpha$\slash Fe]}\,$\sim$\,0.2) than the main body of the host galaxy, supporting a prolonged star formation history~\citep{2010A&A...518A..32M}.

VCC 2095 (NGC 4762). This edge-on lenticular galaxy shows an unusual internal structure that consists of four distinct components~\citep{1984ApJS...56..283W,1989ApJ...339..783H,2004ARA&A..42..603K}. \citet{1984ApJS...56..283W} presumed that the four components are the bulge, bar, lens, and outer ring.
The galaxy shows an extremely small bulge-to-total light ratio ($B/T=0.13 \pm 0.02$) as a S0 galaxy.
\citet{2012ApJS..198....2K} suggested that this late-type S0 galaxy bridges the gap between S0 and spheroidal galaxies.

VCC 698 (NGC 4352). This lenticular galaxy has both small-scale and large-scale stellar disks~\citep{2006ApJS..164..334F}.
Based on the spectroscopic data, the bulge appears to be younger and more metal-rich than the outer part~\citep{2014MNRAS.441..333J}.

VCC 1938 (NGC 4638). This edge-on lenticular galaxy is a disk-dominated system~\citep{2007ApJ...661L..37B}.
\citet{2012ApJS..198....2K} viewed it as a ``missing link'' between S0s and spheroidals like VCC 2095.
They suggested that the galaxy was created from a late-type galaxy by dynamical heating due to surrounding galaxies.

VCC 1883 (NGC 4612). This lenticular galaxy has a bar, a ring, and an isophotal twist. The bar shows misalignment with the main body of the galaxy~\citep{2006ApJS..164..334F}.

VCC 778 (NGC 4377). This lenticular galaxy shows an isophotal twist due to the misaligned bar. There are three spiral galaxies that are regarded as a background group~\citep{2006ApJS..164..334F}.
Thus, there is a possibility that its GC color distribution is contaminated by GCs in the background galaxies.

VCC 1087 (IC 3381). The luminosity-weighted age of this dwarf elliptical galaxy is estimated to be 5\,$\sim$\,7 Gyr~\citep{2006AJ....131..814B,2008MNRAS.385.1374M,2012A&A...548A..78T}.

VCC 1619 (NGC 4550). This lenticular galaxy harbors two counter-rotating disks~\citep{2013A&A...549A...3C} and shows emission lines~\citep{2006MNRAS.366.1151S}. The galaxy also has a peculiar dust distribution that looks like a spiral arm structure~\citep{2001A&A...375..797W}.

VCC 1125 (NGC 4452). This is an edge-on lenticular galaxy. 
\citet{2012ApJS..198....2K} showed that the galaxy has an extremely low pseudo-bulge-to-total luminosity ratio ($=0.017\pm0.004$) and a warped outer disk that seems to stem from gravitational encounters.
They also showed that the galaxy consists of five stellar components like VCC 2095.
The luminosity-weighted mean age using higher-order Balmer absorption lines is $\sim$\,5 Gyr~\citep{2003AJ....125.2891C}.

VCC 1910 (IC 809). This dE or Sph,N type galaxy~\citep{2009ApJS..182..216K} shows a possible disk feature~\citep{2006AJ....132..497L}. The luminosity-weighted mean age based on spectroscopy is 7.5\,$\sim$\,9 Gyr~\citep{2008MNRAS.385.1374M,2014ApJ...783..120T}.

FCC 21 (NGC 1316). This is the only galaxy for which our coeval model does not mimic the CDF. The galaxy is lenticular and well known for its recent merger feature~\citep{1980ApJ...237..303S,2001A&A...376..837H} and the presence of intermediate-age GCs~\citep{2001MNRAS.322..643G,2018MNRAS.479..478S}.
The tidal tails, loops, and arms are extended out to $54\arcsec$ and lots of dust structures are visible in the central region~\citep{1980ApJ...237..303S}.

FCC 276 (NGC 1427). This elliptical galaxy has a kinematically decoupled core in the central region ($\sim$\,7$\arcsec$)~\citep{2014MNRAS.441..274S}.

NGC 1340 (a.k.a. NGC 1344). This elliptical galaxy shows a clear shell structure in the $V$-band image~\citep{2013ApJ...768L..28H}.

FCC 249 (NGC 1419). This E0 galaxy has anomalously bright $K_{S}$-band surface brightness fluctuation~\citep{2002ApJ...564..216L}, which can be induced by recent star formation.

Among 24 galaxies with $t\textsubscript{best-fit}<11$ Gyr plus FCC 21, seven galaxies (VCC 944, VCC 1146, VCC 1913, VCC 2092, FCC 47, FCC 63, and IC 2006) lack the observational data in the literature needed to grasp the characteristics of their morphologies or stellar populations.

\begin{figure*}
\begin{center}
\includegraphics[width=12.0cm]{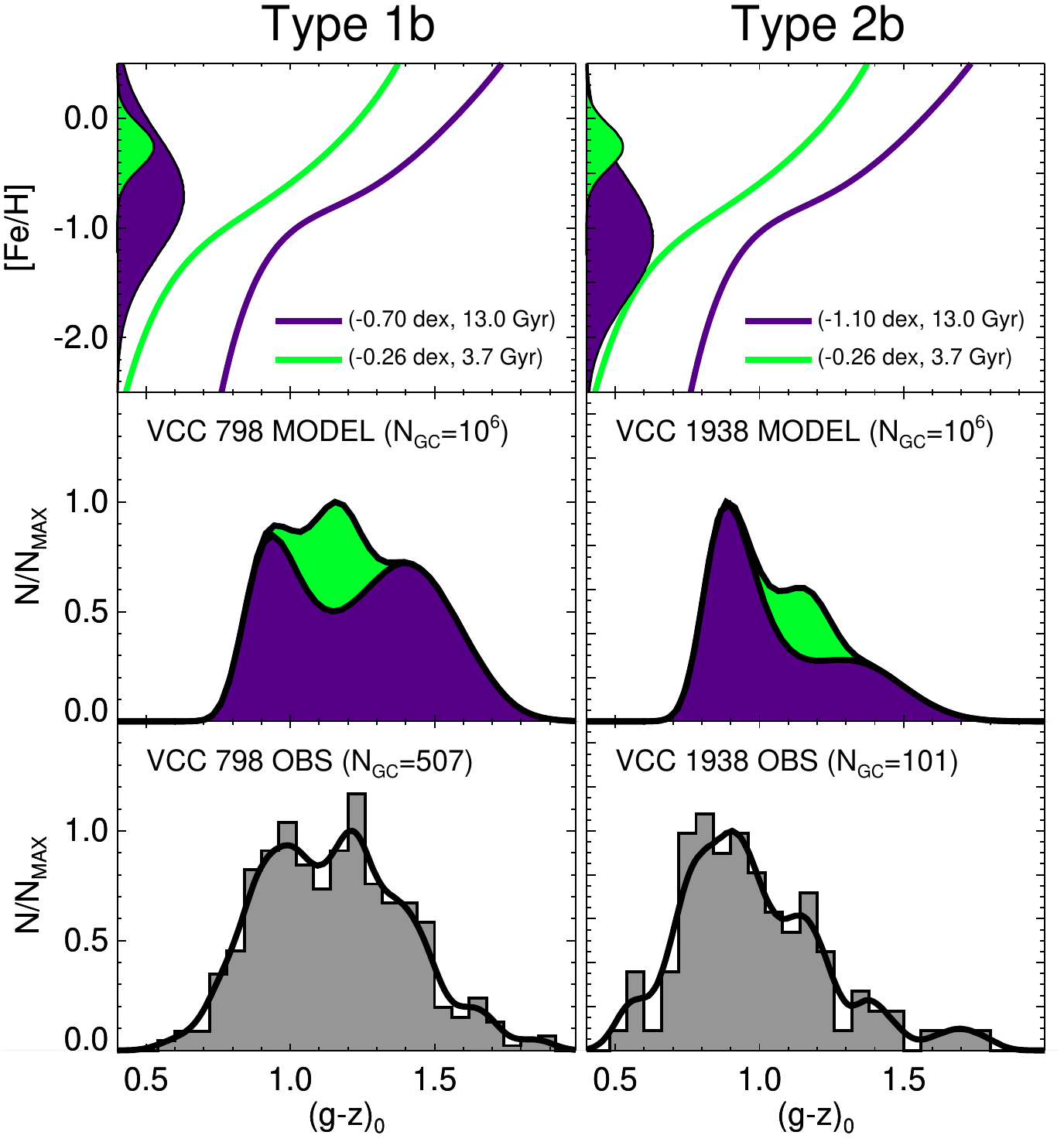}   
\caption{Two-component model simulations for the GC systems with representative CDFs of Types 1b and 2b, VCC 798 and VCC 1938, respectively. Top panels: purple and green solid lines show the ($g-z$)--[Fe/H] relations of the first (old) and the second (young) GC populations, respectively. The assumed GC metallicity distributions are presented by two Gaussian functions on the y-axis. $\sigma$([Fe/H]) of the first GC populations of both VCC 798 and VCC 1938 is assumed to be 0.55. $\sigma$([Fe/H]) of the second GC populations is assumed to 0.2 for both galaxies. The number fraction of the second GC populations is set to be 20\,\% of the first GC populations. The mean metallicities and ages used in the simulations are denoted in parentheses. Middle panels: the simulated GC CDFs; the purple and green histograms correspond to the first and second GC populations, respectively. Bottom panels: the observed GC CDFs of VCC 798 and VCC 1938. The Gaussian kernel density estimations with $\sigma(g-z)=0.06$ are shown by black solid lines.
\label{fig6}}
\end{center}
\end{figure*}

\subsection{Two-component Model for the GC Systems with $t\textsubscript{best-fit}<11$ Gyr}
\label{two-componentmodelfortheglobularclustersystemswitht_best-fit11gyr}

Given that most galaxies with $t\textsubscript{best-fit}<11$ Gyr (namely, Type 1b and 2b galaxies) seem to have experienced recent mergers and\slash or star formation, we hypothesize that their ``diluted'' bimodality is due to contamination of old GC CDFs (i.e., Types 1a and 2a) by young (or intermediate-age) GCs.
We further surmise that Types 1b and 2b are respectively nothing but Types 1a and 2a with additional young GCs.
In order to test whether additional younger GC populations are the origin of Types 1b and 2b CDFs, we have performed two-component (old GCs + young GCs) model simulations.

Figure 6 shows the simulated CDFs for VCC 798 and VCC 1938, the representative examples of Types 1b and 2b, respectively.
As mentioned in Section 5.2, \citet{2018ApJ...859..108K} estimated the spectroscopic ages and metallicities of 20 GCs in VCC 798 (M85).
Out of 20 GCs, 11 are classified as young GCs and their mean age is measured to be $3.7\pm1.9$ Gyr with ${\langle{\rm}}${[Fe\slash H]}${\rangle}$ = $-0.26$.
We adopt these values in modeling the young GCs of VCC 798, as well as VCC 1938 (due to lack of spectroscopic data on VCC 1938 GCs).
The young GCs on top of the old GCs tend to weaken bimodality (of Type 1a) and reduce the sharpness of blue peaks (of Type 2a), bringing the theoretical predictions in good agreement with the observations.
It is highly likely that Types 1b and 2b are the results of adding a young or intermediate-age GC population to Types 1a and 2a CDFs, respectively.
For better assessment of young or intermediate-age GC populations, further spectroscopic observations for the GC systems are badly needed.

\section{D\MakeLowercase{iscussion}}
\label{discussion}

Among 78 early-type galaxies (with $N\textsubscript{GC}$\,$>$\,50) in ACSVCS and ACSFCS, 53 galaxies ($\sim$\,70\,$\%$) have GC color distributions that can be well reproduced by old ($t \textsubscript{best-fit}>11$ Gyr) and coeval model CDFs (Types 1a and 2a).
We have shown that the bimodal color distributions of Types 1a and 2a are attributed to the nonlinear metallicity-to-color conversion of GCs.
The difference in Types 1a and 2a is simply due to the difference in the mean GC metallicity, in that Type 1a is more metal-rich than Type 2a.
On the other hand, our experiment has suggested that for $\sim$\,30\,$\%$ of the sample, young or intermediate-age GCs contaminate the color distribution of underlying old GCs, diluting the color bimodality (of Type 1a) or broadening blue peaks (of Type 2a). We suggest that Types 1b and 2b are respectively nothing but Types 1a and 2a with additional young or intermediate-age GCs.

Our simulations show that for most Types 1b and 2b galaxies, young or intermediate-age GCs in a galaxy comprise $\sim$\,20\,$\%$ of underlying old GCs in number.
This is $\sim$\,17\,$\%$ (= 20/120) with respect to the total number of (old + young) GCs. If $\sim$\,30\,$\%$ of galaxies host $\sim$\,17\,$\%$ of young GCs, the number fraction of the young GCs is as low as $\sim$\,5\,$\%$ of the total GCs of the entire galaxy sample.
This is why the ACSVCS CDFs for GCs in the seven bins of host galaxy luminosity~\citep{2006ApJ...639...95P} are successfully reproduced under the assumption of no young GCs but old, coeval GCs (see Figure 3 in Paper I).

Our methodology usually classifies CDFs as Type 1b or 2b if there are additional young or intermediate-age GCs comprising more than $\sim$\,10\,$\%$ of old GCs.
For additional GCs less than $\sim$\,10\,$\%$, our classification scheme is less reliable. 
For instance, VCC 731 (NGC 4365), which is reported to host an additional GC population~\citep{2003ApJ...585..767L,2005AJ....129.2643B,2005ApJ...634L..41K,2012MNRAS.420...37B}, is classified as Type 1a in our model (Table 2). 
The observed CDF shows that its dip is slightly shallower than other Type 1a galaxies. 
The best-fit $\sigma(${[Fe\slash H]}$)$ and age of this galaxy are 0.45 and 12.5 Gyr, respectively, but if $\sigma(${[Fe\slash H]}$)$ is set to $>0.5$, K-S statistics prefer a younger ($<11$ Gyr) CDF model (see Figure 7).
This phenomenon is caused by the fact that both the smaller $\sigma(${[Fe\slash H]}$)$ and the fill-in of additional GCs alike make the dip of a CDF shallower.
Thus, some Type 1a galaxies whose best-fit $\sigma(${[Fe\slash H]}$)$ values are smaller than 0.5 might contain some ($<$\,10\,$\%$) young or intermediate-age GCs.

There exist some GC systems that exhibit CDFs with two strong peaks with only a few (if any) GCs in between. 
Good examples are NGC 3115~\citep{2012ApJ...759L..33B} in isolation and NGC 1387 and NGC 1404 near NGC 1399 in the Fornax cluster~\citep{2013ApJ...763...40K}. Their CDFs are also characterized by red GCs that are more abundant than other galaxies with similar luminosity.
Our sample contains some galaxies showing an unusually high red GC fraction for their luminosities and ${\langle{\rm}}${[Fe\slash H]}${\rangle}$ (e.g., VCC 1146, FCC 167, FCC 184 = NGC 1387, and FCC 219 = NGC 1404).
The strong bimodality with few GCs in between and/or the unusually high red GC fraction may point to metal-rich GCs created by heavy star formation after metal-poor GCs were generated earlier on.
This seems consistent with the prediction by the two-phase galaxy formation scenario, in that in low-mass galaxies, star formation is often prolonged at a significant level toward lower redshift~\citep[see Figure 6  in][]{2010ApJ...725.2312O}.

Even with the upcoming large telescopes, it is extremely challenging for spectroscopic observations to quantify the number of young or intermediate-age GCs with respect to underlying, old GCs.
We hence anticipate that systematic analyses of the ``dilution'' effect of GC CDFs, equipped with precise knowledge on the shape of the GC CMRs, will become a vital tool for examining the star formation histories of external galaxies beyond the Virgo and Fornax clusters of galaxies.

\section{C\MakeLowercase{onclusion}}
\label{conclusion}

We have simulated individual $g-z$ CDFs for GC systems of 78 early-type galaxies in Virgo and Fornax clusters using the YEPS model~\citep{2013ApJS..204....3C}. The GC CDFs are from the GC catalogs by ACSVCS~\citep{2009ApJS..180...54J} and ACSFCS~\citep{2015ApJS..221...13J}.
The main results of this study are as follows:

\begin{enumerate}
\item The \textit{HST} $g-z$ color distributions of the majority ($\sim$\,70\,$\%$) of GC systems are naturally reproduced by the nonlinear metallicity-to-color conversion under the simple assumption of the presence of old ($>11$ Gyr), coeval GCs. We refer to them as Types 1a and 2a. The variation of the GC CDFs stems from systematic differences in the mean metallicity of GC systems, in that more luminous galaxies (mostly Type 1a) possess more metal-rich GC systems than less luminous galaxies (mostly Type 2a).

\item The other GC systems ($\sim$\,30\,$\%$) show CDFs that can be best reproduced by the two-component (old GCs + young GCs) model. We refer to them as Types 1b and 2b. Most of the galaxies which host GC systems with $t\textsubscript{best-fit}<11$ Gyr show signs of mergers and/or star formation. These galaxies most likely have additional young or intermediate-age GCs, which alter the color distributions of underlying, old GCs. Types 1b and 2b are suggested to be the variants of Type 1a and 2a, respectively.

\item There is a strong, positive correlation between the mean metallicity and the host galaxy luminosity, as also reported by other studies~\citep[e.g.,][]{2006ApJ...639...95P}. The best-fit age and metallicity dispersion of GCs show no obvious correlation with the host galaxy luminosity.

\end{enumerate}

\acknowledgments
S.J.Y. acknowledges support from the NRF of Korea to the Center for Galaxy Evolution Research (No. 2017R1A5A1070354). This work was partially supported by the KASI--Yonsei Joint Research Program (2018).

\appendix
This Appendix presents the impact of differing $\sigma$([Fe/H]) on the determination of the best-fit age and ${\langle{\rm}}$[Fe/H]${\rangle}$.
Figure 7 shows the $p$-value contour plots for the whole galaxy sample.

\begin{figure*}[h]
\plotone{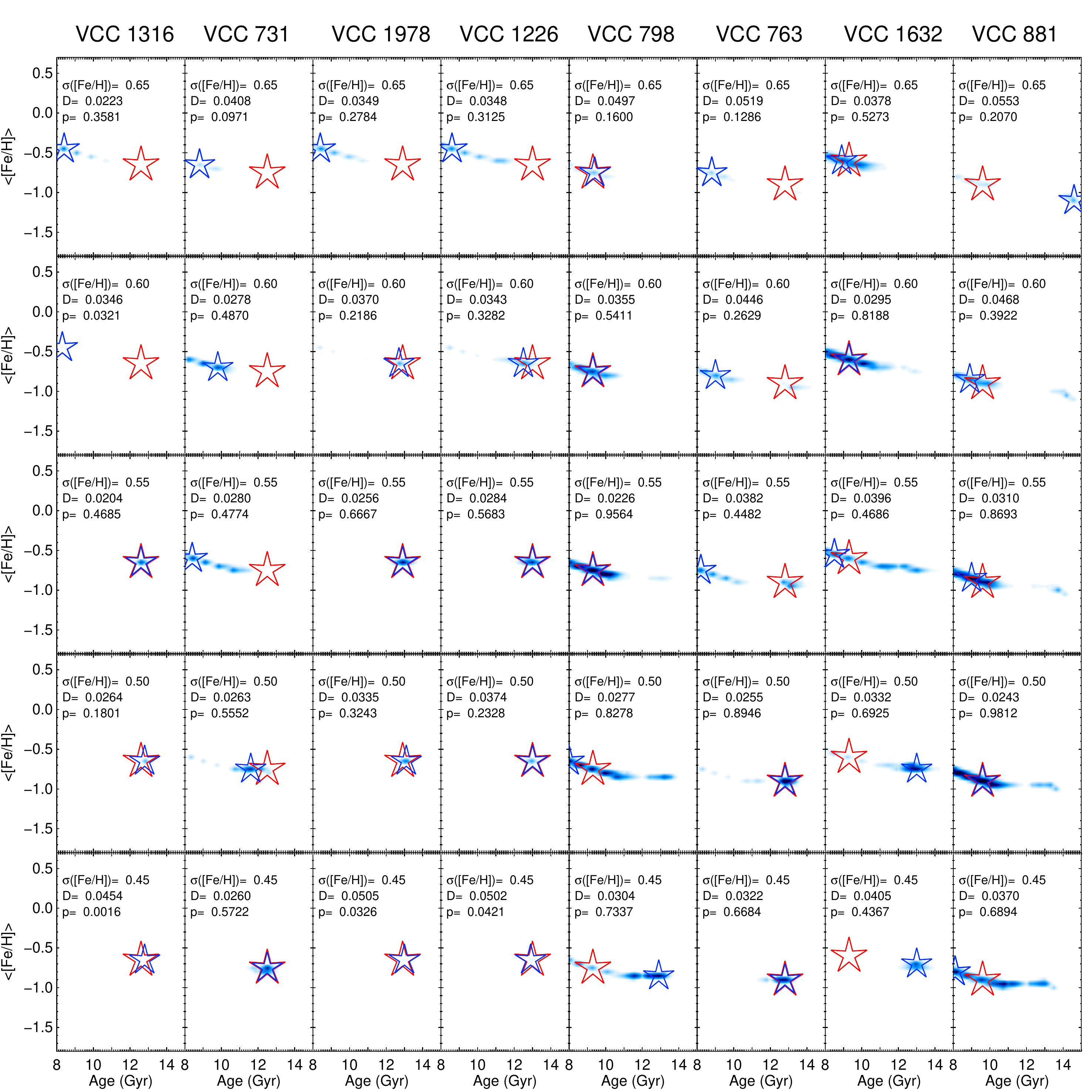}
	\caption{The effect of differing $\sigma$([Fe/H]). Each column presents the confidence contours for age and ${\langle{\rm}}$[Fe/H]${\rangle}$ based on the K-S test between models and observations for each galaxy. Each row shows the results for $\sigma$([Fe/H]) being 0.65, 0.60, 0.55, 0.50, and 0.45 dex (from top to bottom). The quantities of $D$ and $p$-value for each case of $\sigma$([Fe/H]) are also labeled. The contour levels are the same as those in Figure 1. The blue stars show the best-fit age and ${\langle{\rm}}$[Fe/H]${\rangle}$ for each case of $\sigma$([Fe/H]). The red stars represent the best-fit age and ${\langle{\rm}}$[Fe/H]${\rangle}$ for the best-fit $\sigma$([Fe/H]).}
\label{fig7}
\end{figure*}

\begin{figure*}
	\figurenum{7}
	\plotone{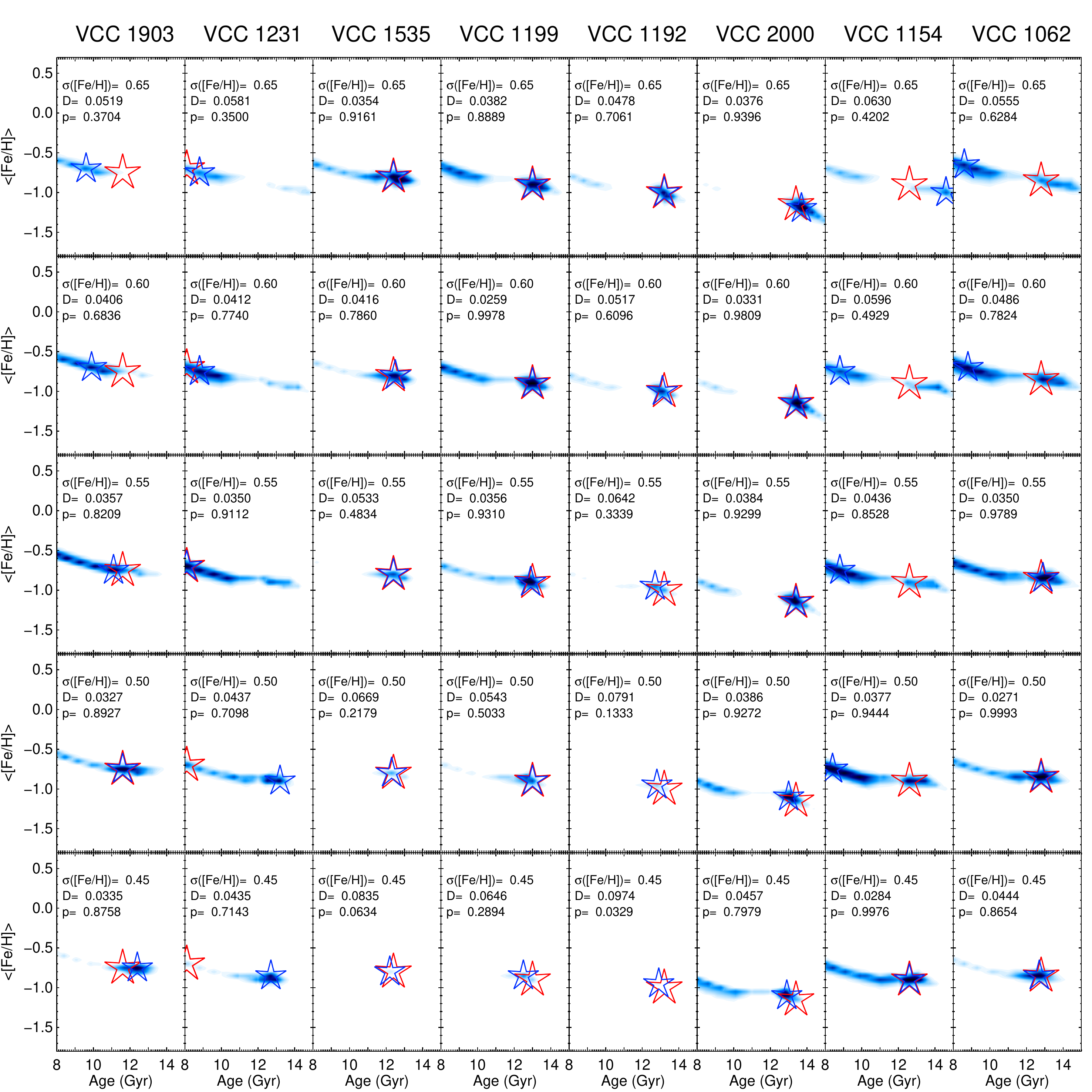}
	\caption{\emph{Continued}}
\end{figure*}
\begin{figure*}
	\figurenum{7}
	\plotone{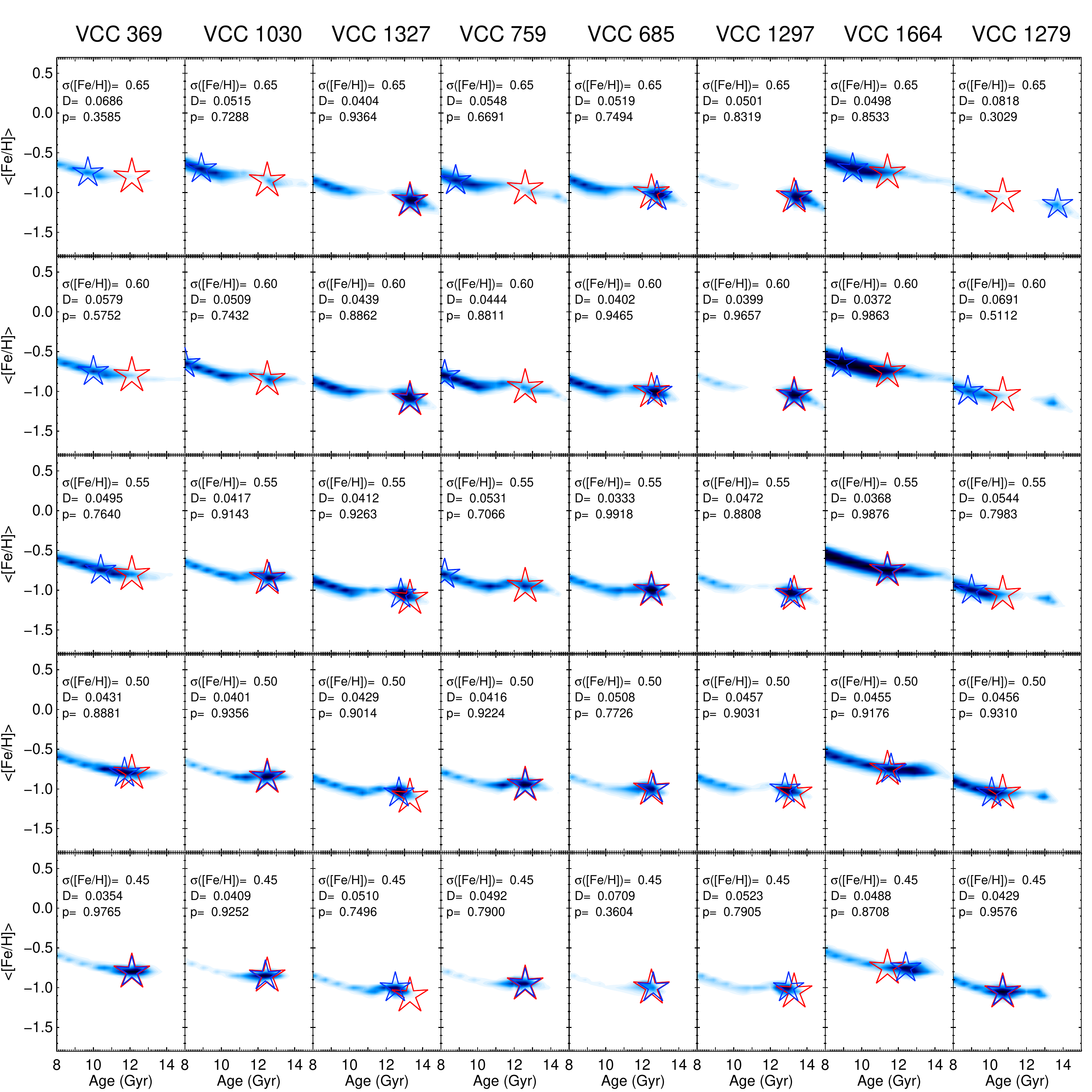}
	\caption{\emph{Continued}}
\end{figure*}
\begin{figure*}
	\figurenum{7}
		\plotone{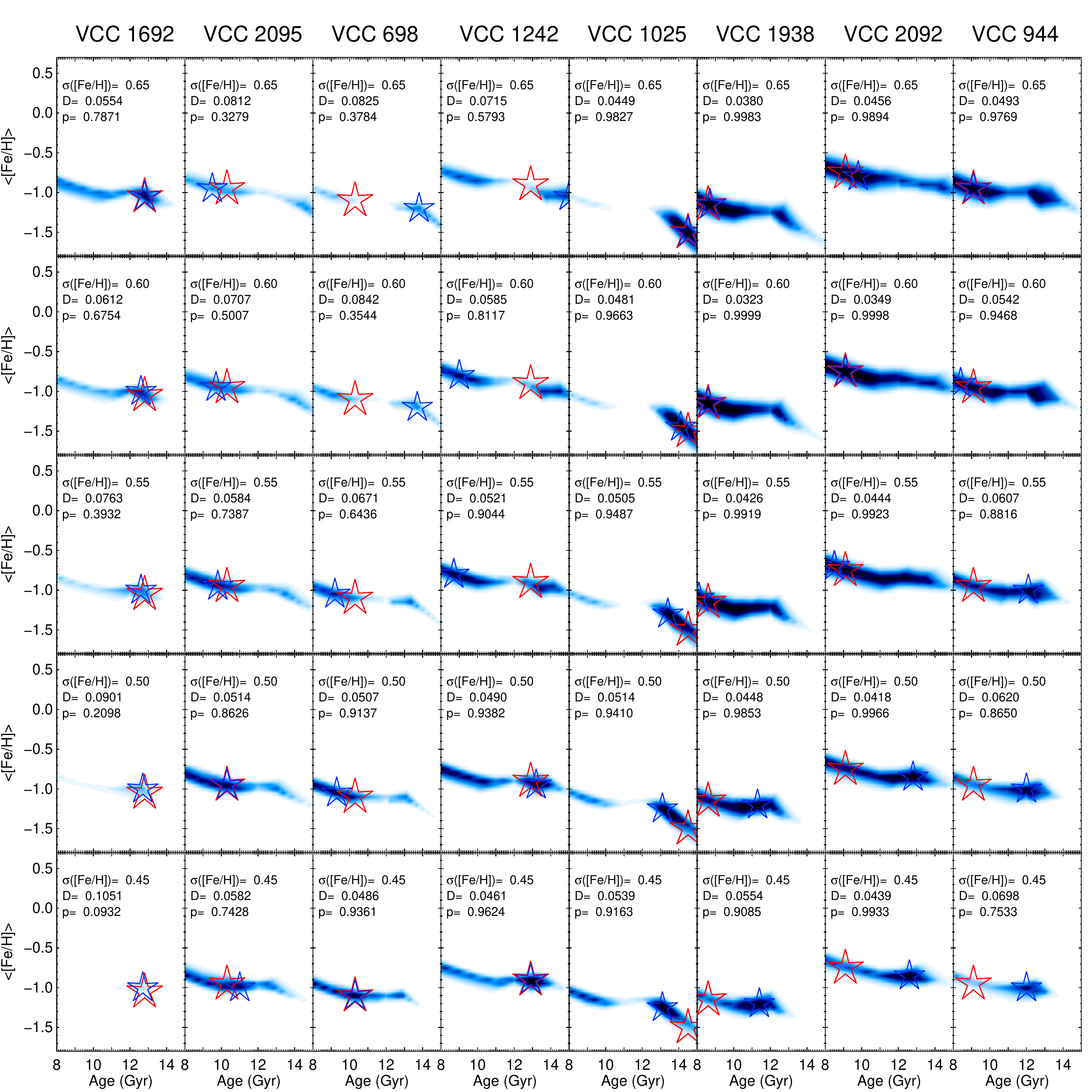}
	\caption{\emph{Continued}}
\end{figure*}
\begin{figure*}
	\figurenum{7}
\plotone{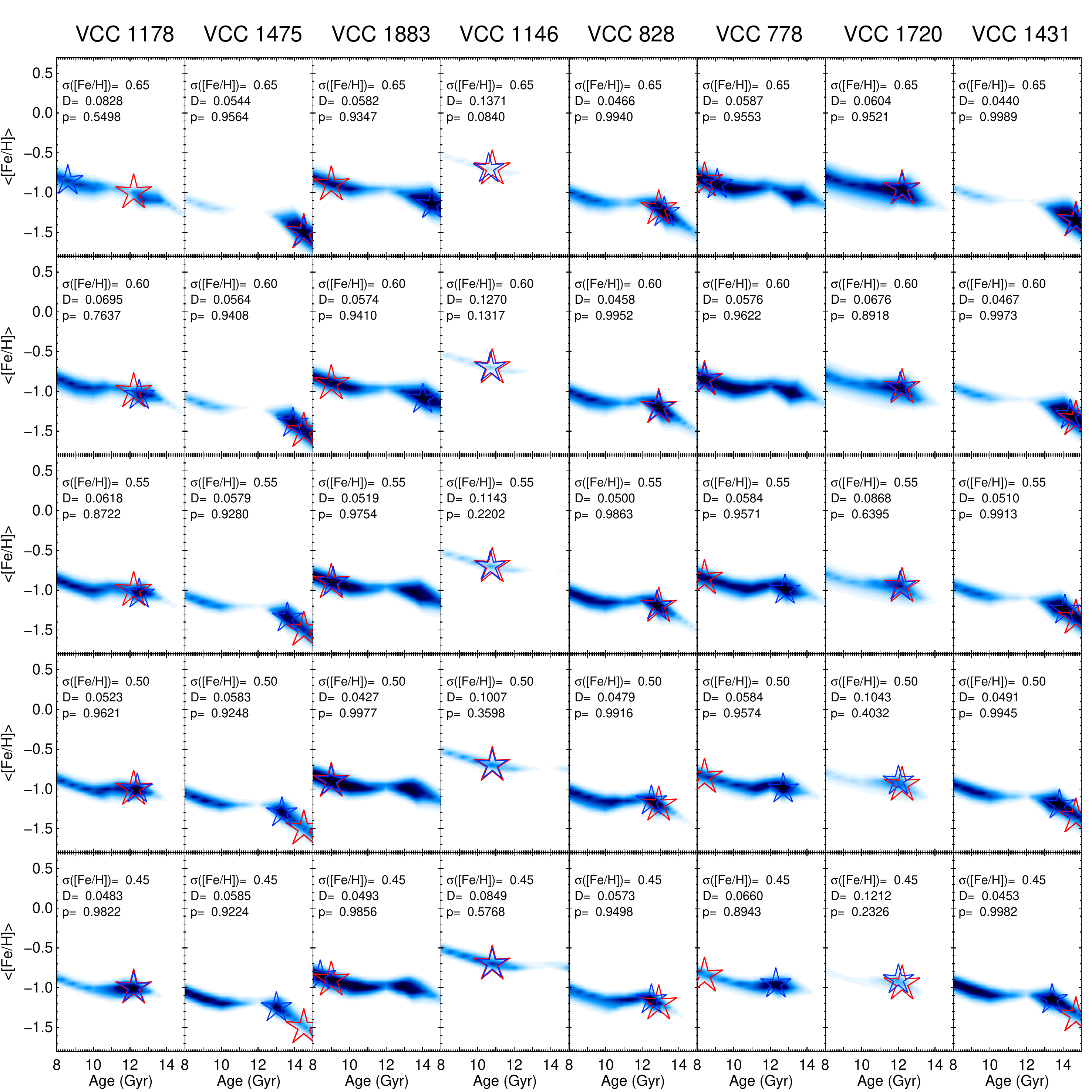}
	\caption{\emph{Continued}}
\end{figure*}
\begin{figure*}
	\figurenum{7}
	\plotone{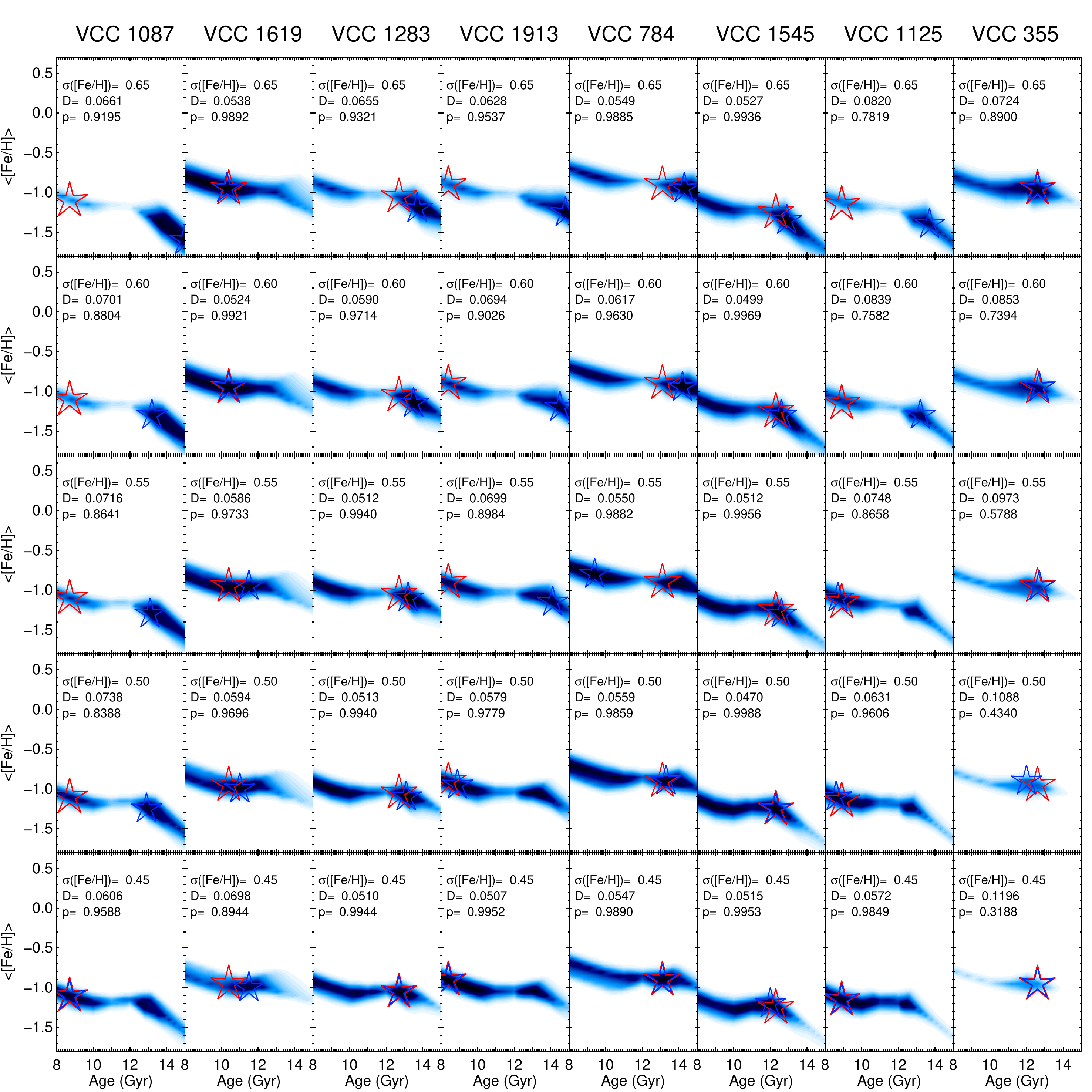}
	\caption{\emph{Continued}}
\end{figure*}
\clearpage
\begin{figure*}
	\figurenum{7}
	\plotone{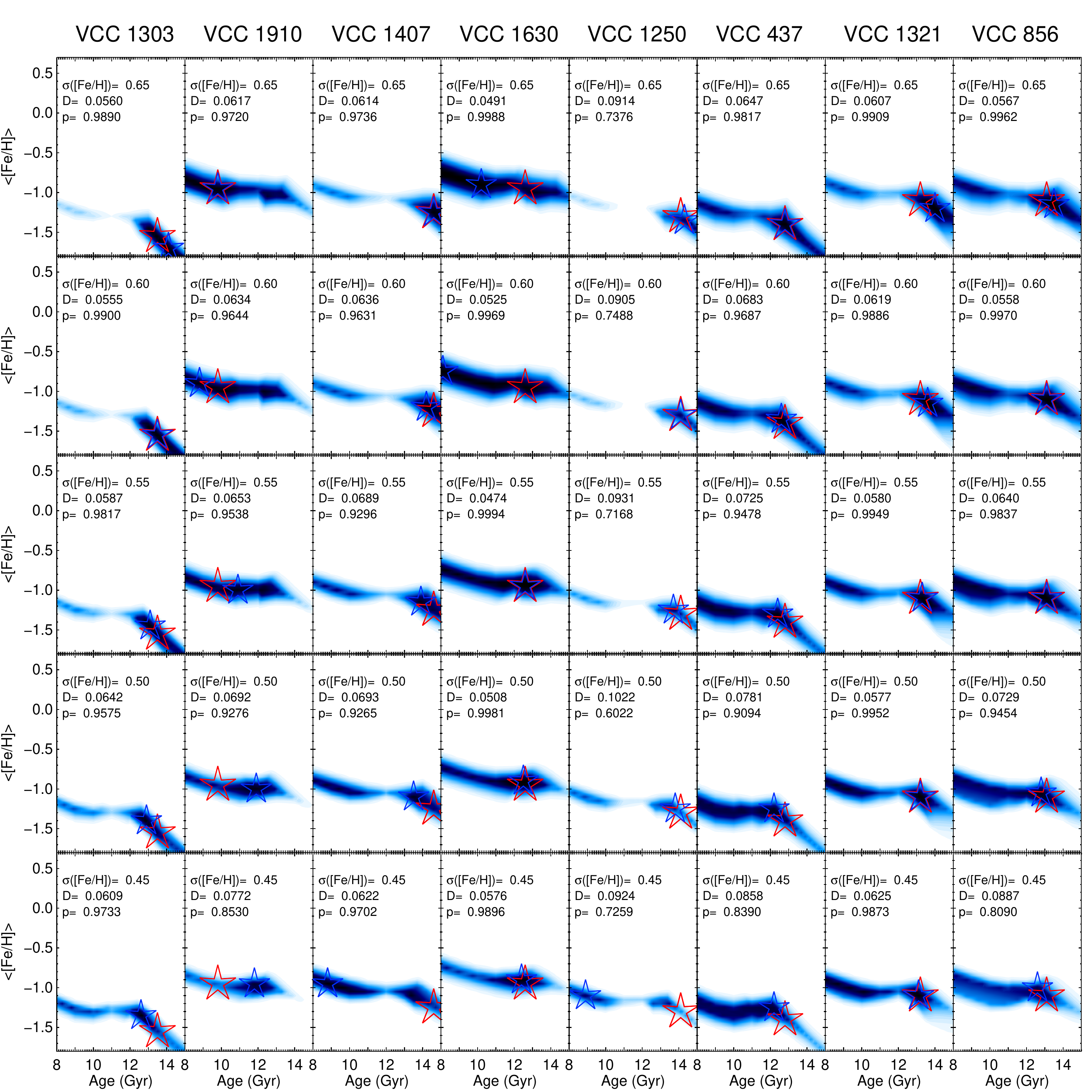}
	\caption{\emph{Continued}}
\end{figure*}
\begin{figure*}
	\figurenum{7}
	\plotone{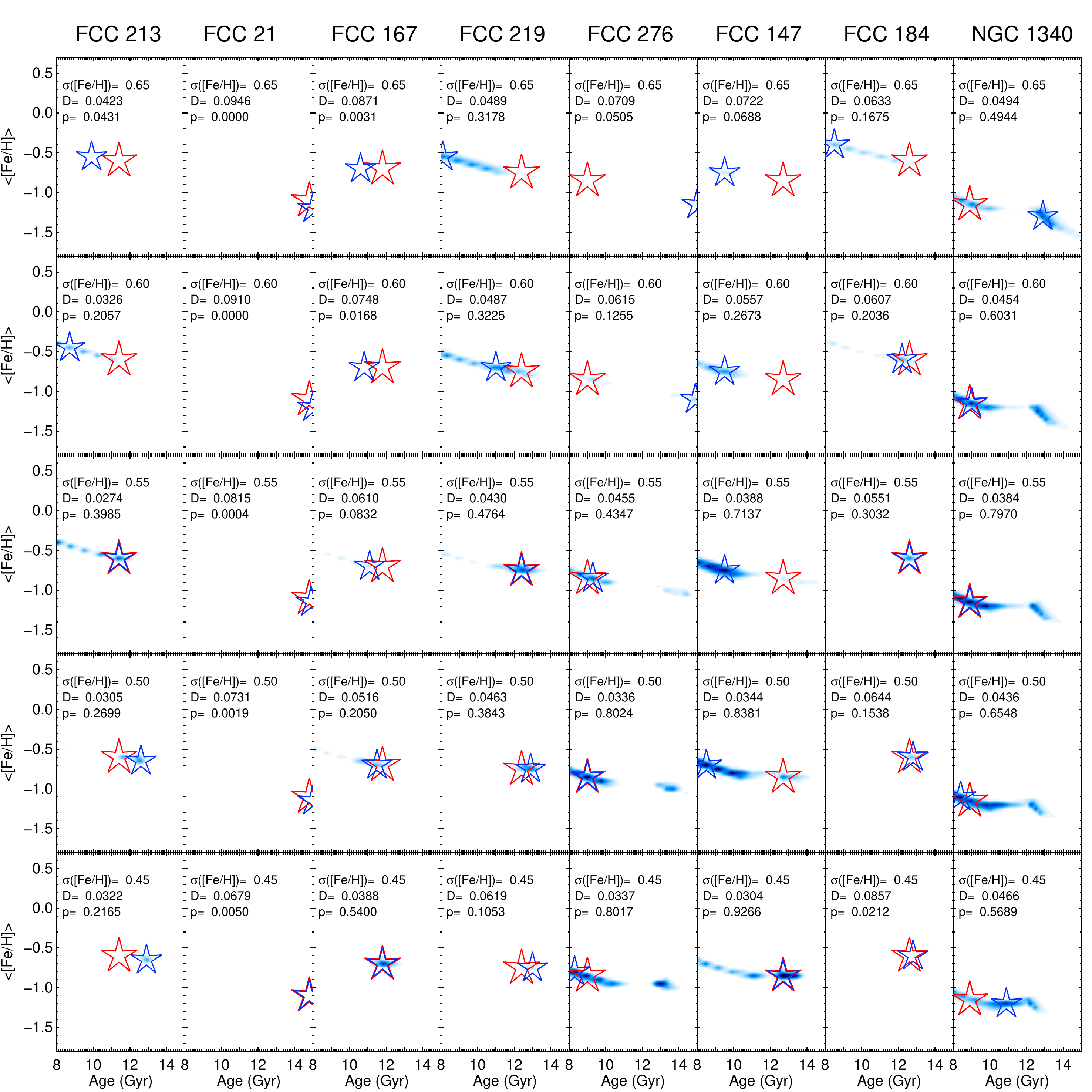}
	\caption{\emph{Continued}}
\end{figure*}
\begin{figure*}
	\figurenum{7}
	\plotone{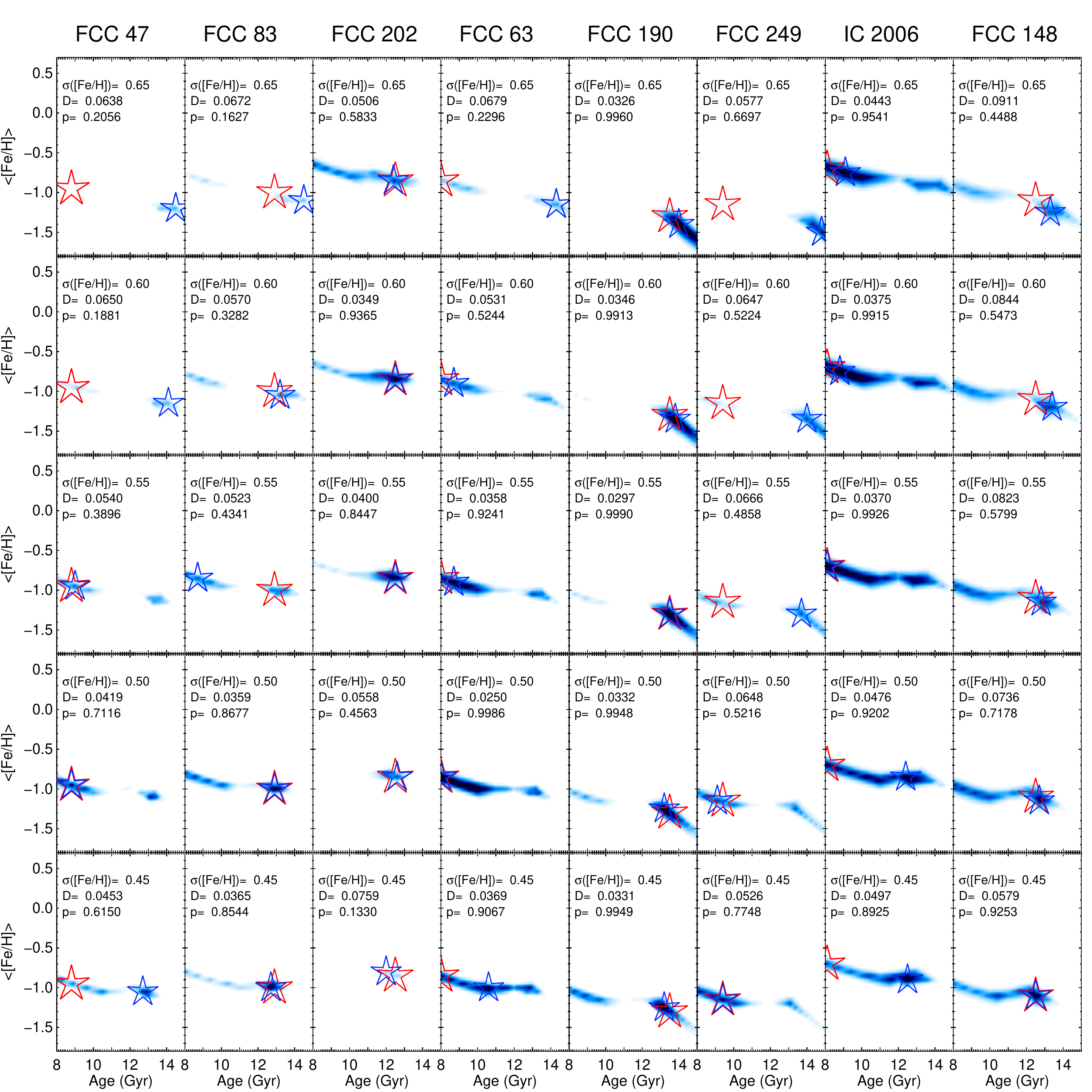}
	\caption{\emph{Continued}}
\end{figure*}
\clearpage
\begin{figure*}
	\figurenum{7}
	\plotone{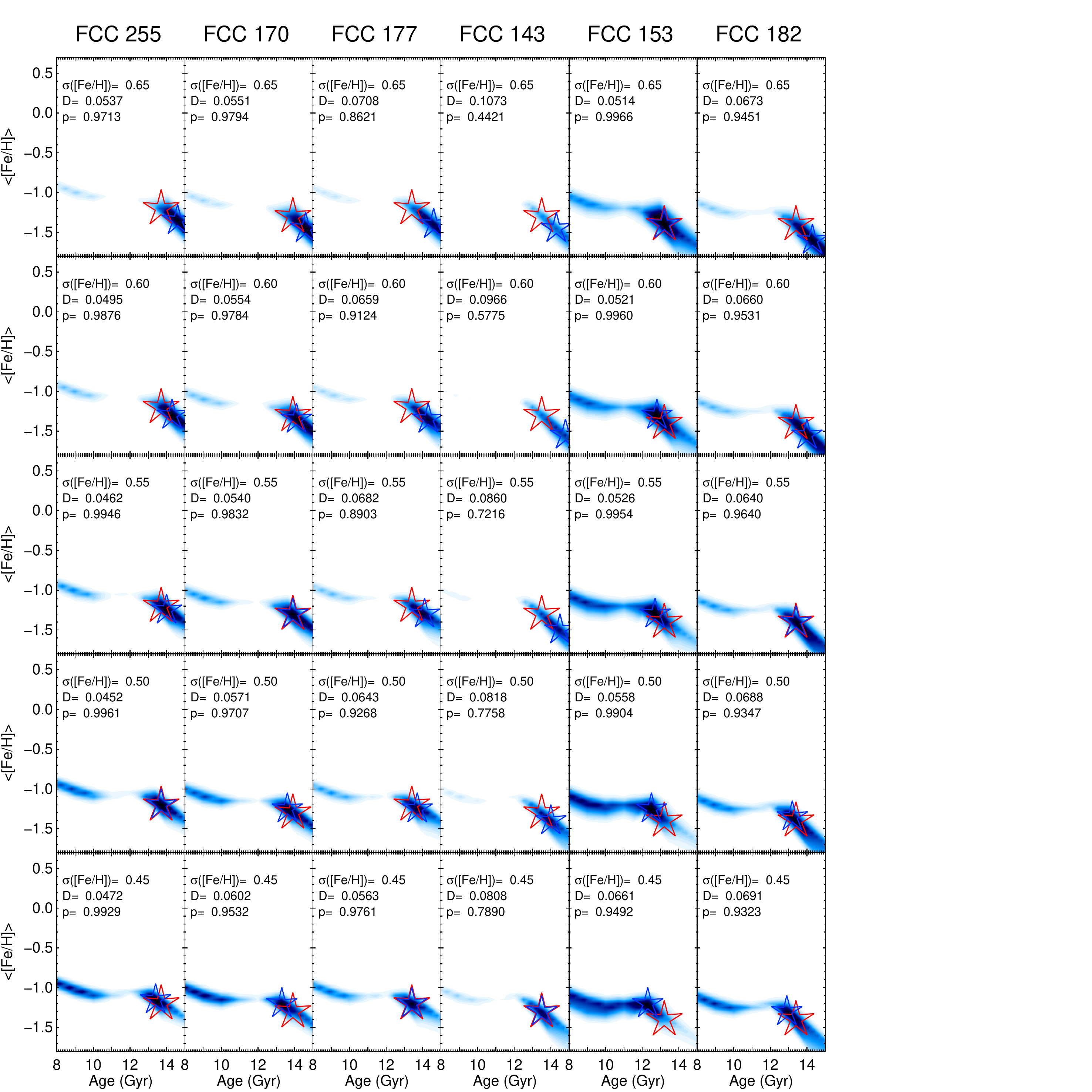}
	\caption{\emph{Continued}}
\end{figure*}

\end{document}